%% file: ms.tex
\pdfoutput=1
\documentclass[english]{paper}

\input{header}

\begin{document}

\include{article}
\include{appendix}

\end{document}

%% file: header.tex
    \usepackage[T1]{fontenc}
    \usepackage[utf8]{inputenc}
    \usepackage{color}
    \usepackage{babel}
    \usepackage{verbatim}
    \usepackage{nicefrac}
    \usepackage{amsmath}
    \usepackage{graphicx}
    \usepackage{csquotes}
    \usepackage{siunitx}
    \usepackage{xr}
    \usepackage{float}

    \usepackage{tikz}
    \usetikzlibrary{tikzmark}
    \usetikzlibrary{calc}

    \usepackage{lineno}

    \usepackage[
        unicode=true,
        pdfusetitle,
        bookmarks=true,
        bookmarksnumbered=false,
        bookmarksopen=false,
        breaklinks=false,pdfborder={0 0 0},
        pdfborderstyle={},
        backref=false,
        colorlinks=false
    ]{hyperref}
    \usepackage[noabbrev,capitalise]{cleveref}

    \usepackage{authblk}

    \usepackage[style=numeric,sorting=none,maxbibnames=99]{biblatex}
    \makeatletter

    \def\inputval{0}
    \newread\inputFile
    \newcommand*{\qtyincludevalue}[3][]{%
      \IfFileExists{./data/output/data-values/#2}{
        \openin\inputFile=./data/output/data-values/#2
        \read\inputFile to \inputval
        \closein\inputFile
        \qty[#1]{\inputval}{#3}%
      }{\qty[#1]{#2}{#3}}%
    }

    \newcommand{\pout}{p_{\mathrm{out}_l}}
    \newcommand{\pin}{p_{\mathrm{in}_l}}
    \newcommand{\Pout}{P_{\mathrm{out}}}
    \newcommand{\Pin}{P_{\mathrm{in}}}

    \newcommand*{\addFileDependency}[1]{%
        \typeout{(#1)}
        \@addtofilelist{#1}
        \IfFileExists{#1}{}{\typeout{No file #1.}}
    }

    \newenvironment{lyxlist}[1]
        {\begin{list}{}
            {\settowidth{\labelwidth}{#1}
             \setlength{\leftmargin}{\labelwidth}
             \addtolength{\leftmargin}{\labelsep}
             }}
        {\end{list}}

    \graphicspath{{./}}

    \makeatother

%% file: article.tex
%auto-ignore
\author[1]{Peter Regner \thanks{Corresponding author: peter-erlwhu@suuf.cc} }
\author[1]{Katharina Gruber }
\author[1]{Sebastian Wehrle }
\author[1]{Johannes Schmidt }

\affil[1]{
    Institute for Sustainable Economic Development,
    University of Natural Resources and Life Sciences, Vienna
}

\title{Explaining the decline of US wind output power density}

\maketitle

\begin{abstract}
US wind power generation has grown significantly over the last decades, in line with the number and average size of operating turbines. However, wind power density has declined, both measured in terms of wind power output per rotor swept area as well as per spacing area. To study this effect, we present a decomposition of US wind power generation data for the period 2001--2021 and examine how changes in input power density and system efficiency affected output power density. Here, input power density refers to the amount of wind available to turbines, system efficiency refers to the share of power in the wind flowing through rotor swept areas which is converted to electricity and output power density refers to the amount of wind power generated per rotor swept area. We show that, while power input available to turbines has increased in the period 2001--2021, system efficiency has decreased. In total, this has caused a decline in output power density in the last 10 years, explaining higher land-use requirements. The decrease in system efficiency is linked to the decrease in specific power, i.e. the ratio between the nameplate capacity of a turbine and its rotor swept area. Furthermore, we show that the wind available to turbines has increased substantially due to increases in the average hub height of turbines since 2001. However, site quality has slightly decreased in this period.

\end{abstract}

\section{Introduction\label{sec:introduction}}

\begin{figure}
        \centering{}\includegraphics[width=\textwidth]{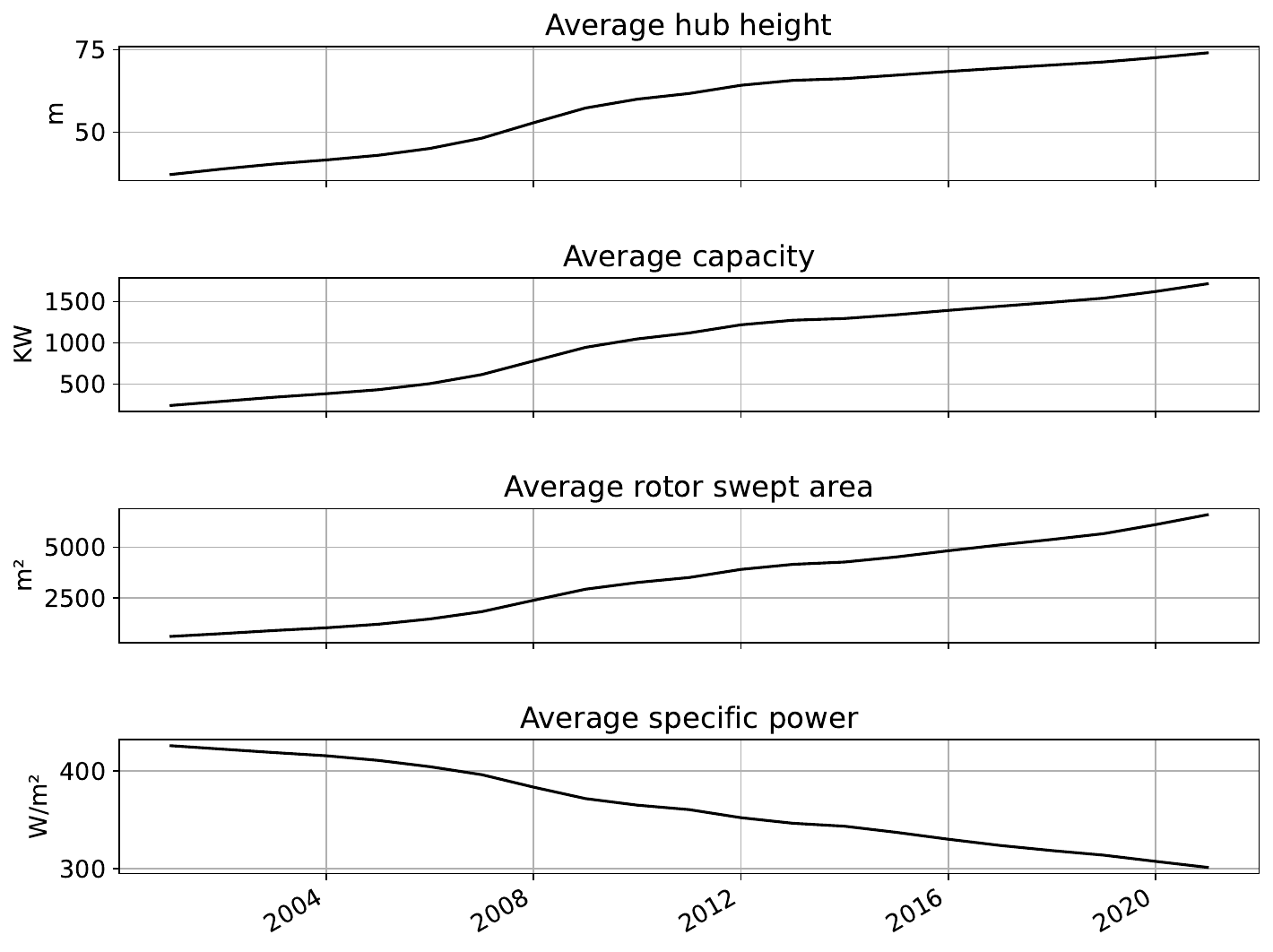}
        \caption{Evolution of turbine characteristics: average hub height, capacity and rotor swept area of operating wind turbine models increases over time, but not at the same pace. Specific power, the ratio between capacity and rotor swept area, shows a declining trend. (Data source: USWTDB, see \cref{sec:methods-data-turbines})
        \label{fig:growth_and_specific_power}}
    \end{figure}

    Wind power generation has been expanding rapidly in the US in the past two decades. Between 2001 and 2021, the number of operating turbines increased fivefold (see \cref{sec:results}). At the same time, new wind turbine models with larger hub height, rotor swept area and capacity were deployed (see \cref{fig:growth_and_specific_power}a-c), which also contributed to more wind power being generated.

    However, while total output increased, output power density declined. Output power density can be measured in terms of power output per rotor-swept area or per spacing area. \citeauthor{miller_observation-based_2018} \cite{miller_observation-based_2018,miller_corrigendum:_2019} have shown a decline in power output per spacing area, implying higher land requirements for wind power, while we show here that power output per rotor swept area also declined.

\begin{figure}
        \centering{}\includegraphics[width=\textwidth]{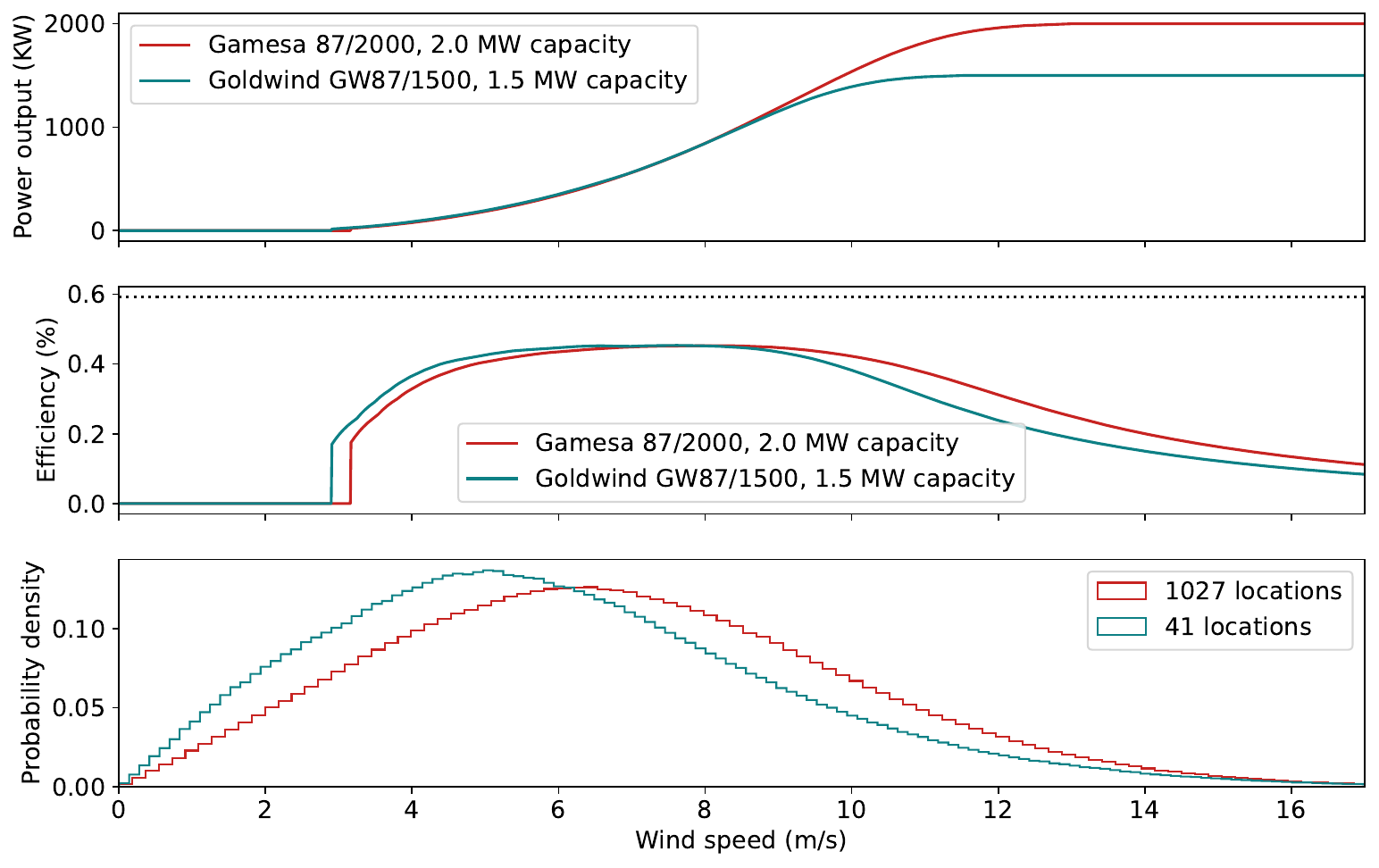}
        \caption{Comparison of two turbine models with \qty{87}{m} rotor diameter: The turbine model \emph{Gamesa 87/2000} has a higher capacity of \qty{2}{MW} compared to \emph{Goldwind GW87/1500} with \qty{1.5}{MW}. Since rotor diameters are equal, the Gamesa model also has a higher specific power (\qty{336.4}{W/m^2}) than the Goldwind model (\qty{252.3}{W/m^2}). At wind speeds above about \qty{8}{m/s}, the Gamesa model has a higher power output. Despite economic advantages of turbine models with lower specific power, higher specific power turbines generate more power per rotor swept area if generated energy can be stored or transmitted. Wind speeds above \qty{8}{m/s} are less likely than lower wind speeds (third panel of the figure), but higher wind speeds play a larger role because wind power depends on the cube of the wind speed. In total, output power density decreases due to the decline of specific power as will be shown in the following. (Data sources: power curve model \cite{ryberg_future_2018}, ERA5 \cite{reanalysis_era5_single_levels} bias-corrected with GWA2 \cite{GWA2}, see also \cref{sec:methods-data}).\label{fig:example_turbine_characteristics}}
    \end{figure}

    We explore the factors which contributed to the decline in output power density. In particular, we assess how input power density, i.e. the wind available to turbines, changed -- due to new location choices, but also due to the increase in average hub heights. Furthermore, we assess the contribution of system efficiency to the decrease in output power density. System efficiency measures how much of the wind available to turbines is converted to electricity. One main factor determining system efficiency is the specific power of turbines, i.e. the ratio of nameplate capacity to rotor swept area: a decrease in the average specific power of turbines will decrease, everything else equal, system efficiency. The relation between specific power and system efficiency is illustrated in \cref{fig:example_turbine_characteristics}. Wind power generation of a single turbine at a certain point in time is capped by its rated capacity. Therefore, turbines with lower specific power inevitably use less of the available wind resources compared to turbines with higher specific power when high wind speeds occur. When comparing turbines with equal rotor diameter but different capacity, the one with higher capacity and therefore higher specific power has higher power output at high wind speeds.

    Specific power has decreased strongly in the US in the recent 20 years (see \cref{fig:growth_and_specific_power}d), because turbines with lower specific power are economically more profitable for operators \cite{bolinger_opportunities_2020} and easier to integrate into the system \cite{HIRTH201651}. This development can explain potential decreases in system efficiency.

    Here, we therefore assess historic trends in output power density, i.e. how much electric power can be generated per rotor swept area, considering the whole US fleet, to deepen our understanding of declining land-use efficiency of wind turbines. Furthermore, we show how the change in output power density is related to: (a) the change in the wind available to the fleet, measured by power in the wind per rotor swept area (i.e. input power density) and (b) how efficiently the fleet converts the power in the wind to electric power (system efficiency). Furthermore, we decompose the change in input power density into change due to new locations, due to increasing average turbine heights and due to annual variations in wind conditions. To derive those indicators, we introduce a novel decomposition approach, which we apply to the complete US fleet in the period 2001--2021. This allows us to answer (a) if more wind is available to wind turbines compared to previous periods, (b) why wind availability changed, and (c) how the efficiency of converting wind to electric power has evolved over time. It also explains why spacing area requirements of wind turbines have increased recently, as rotor swept area and spacing area requirements are intrinsically related.

\section{Definition of wind power metrics\label{sec:wind-power-metrics}}
    Here, we first introduce several metrics to conceptualize our understanding of trends in US wind power generation.
    The yield of wind power generation, i.e. the total
    generated electric power averaged over a certain period of time for all operating wind turbines in the US, will be denoted as \emph{power output} or with the symbol $\Pout$.
    Power output is the part of available wind resources which is converted to electricity. \emph{Power input} $\Pin$ is the primary energy in wind which is harnessed. It is the total kinetic power of moving air flowing through the rotor swept areas of all operating wind turbines in the US, neglecting disturbances introduced by turbines themselves, e.g. wake effects.  For a single wind turbine, power input is proportional to the rotor swept area of the turbine (see \cref{sec:appendix-alternative-definitions-efficiency}). Therefore, at fixed turbine heights, turbine locations can be compared by the power input relative to the given area. Total rotor swept area of operating turbines is denoted by $A$ and \emph{input power density} $\frac{\Pin}{A}$ is the power input normalized by the total rotor swept area. Furthermore, we define \emph{output power density} as the amount of generated electric power per unit of rotor swept area $\frac{\Pout}{A}$. Note that $A$ can be interpreted as the total size of the wind power fleet which is linked to land use requirements and impacts, because both length of turbine blades of turbines and number of operating turbines contribute to it.

    With \emph{system efficiency} we denote the share of power input converted to electricity, i.e. the ratio of power output and power input $\frac{\Pout}{\Pin}$. System efficiency is bound by the Betz' limit and results from different influencing factors, such as mechanical and technical efficiency, but also system effects such as input power density, wakes, curtailing and downtime due to maintenance, as discussed in more detail in \cref{sec:results-efficiency}.

    Output power density can be viewed as a combination of input power density and system efficiency:

    \begin{equation}
        \frac{\Pout}{A} =
            \frac{\Pin}{A} \cdot
            \frac{\Pout}{\Pin}.
        \label{eq:power-output-decomposition}
    \end{equation}

    The introduced terminology allows for a multiplicative decomposition of wind power output into the total number of operating turbines, average rotor swept area per turbine, input power density, and system efficiency. Denoting the total number of operating turbines with $N$, the decomposition can be formally described by

    \bigskip
    \begin{minipage}{0.95\linewidth}
    \begin{equation}
        \Pout =
            \tikzmark{number_of_turbines}N \cdot
            \tikzmark{avg_rotor_swept_area}\frac{A}{N} \cdot
            \tikzmark{input_powerdensity}\frac{\Pin}{A} \cdot
            \tikzmark{efficiency}\frac{\Pout}{\Pin}.
            \label{eq:multiplicative-decomposition}
    \end{equation}
    \begin{tikzpicture}[remember picture,overlay]
        \draw[<-]
          ([shift={(10pt,-14pt)}]pic cs:efficiency) |- ([shift={(14pt,-24pt)}]pic cs:efficiency)
          node[anchor=west] {$\scriptstyle \text{system efficiency}$};
        \draw[<-]
          ([shift={(8pt,-14pt)}]pic cs:input_powerdensity) |- ([shift={(14pt,-36pt)}]pic cs:input_powerdensity)
          node[anchor=west] {$\scriptstyle \text{input power density}$};
        \draw[<-]
          ([shift={(5pt,-14pt)}]pic cs:avg_rotor_swept_area) |- ([shift={(14pt,-48pt)}]pic cs:avg_rotor_swept_area)
          node[anchor=west] {$\scriptstyle \text{average rotor swept area per turbine}$};
        \draw[<-]
          ([shift={(5pt,-14pt)}]pic cs:number_of_turbines) |- ([shift={(14pt,-60pt)}]pic cs:number_of_turbines)
          node[anchor=west] {$\scriptstyle \text{total number of operating turbines}$};
    \end{tikzpicture}
    \vspace*{60pt}
    \end{minipage}

    In \cref{sec:appendix-motivation}, we show how our decomposition is motivated by the physics behind wind to power conversion for deriving wind power generation from wind speeds. For each of the introduced variables, a time series will be calculated for the period 2001--2021. $\Pin$ is computed by estimating the kinetic power in wind at all known US wind turbine locations from reanalysis wind speed data which is bias corrected with the global wind atlas. $\Pout$ uses the wind speed data in combination with a generic power curve model, based on specific power, to derive wind turbine power output for all known turbines in the US (see \cref{sec:methods-computation-p-out}).

    However, there are no precise turbine commissioning dates available, but only the year of installation for each turbine, which is why we cannot calculate time series for $A$ and $N$ with a resolution higher than yearly. Additionally wind speeds are subject to large seasonal variations. We therefore define the variables $\Pin$, $\Pout$, $N$, and $A$ as yearly aggregated time series for all wind turbines in the US and use them for all further computations. That means the time series $\Pin$, $\Pout$, $N$, and $A$ are functions of the year $Y$.

    For $Y$ being the set of all hours in a year and $L$ being the set of all turbine locations, the aggregated time series $\Pin$ is given by
    \begin{equation}
        \Pin \left(Y\right)
            = \frac{1}{\left|Y\right|} \sum_{t\in Y} \sum_{l\in L} b_{Y,l} \cdot \pin \left(v_{t,l}\right),
        \label{eq:p_in-formula}
    \end{equation}
    where $\pin \left(v_{t,l}\right)$ is the kinetic power in wind flowing through the rotor swept area at location $l$ during hour $t$ (see \cref{sec:appendix-motivation}) and $b_{Y,l}$ indicates whether turbine $l$ was already built and operating in year $Y$ (see \cref{sec:methods-computation-p-in}). Due to the coarse resolution of the wind speed data, we assume that wind speed reductions and wake effects caused by the turbines are not reflected in the wind speed data and therefore also not part of $\Pin$. Hence, system efficiency relates generated electric power to the theoretically available wind resources, which can be captured by the turbine.

    Similarly, $\Pout\left(Y\right)$ is the average generated electric power in time period $Y$ of all wind turbines. A power curve model is used to estimate $\pout \left(v_{t,l}\right)$ and a constant loss correction factor is applied to account for downtime, wakes and other losses not reflected in the power curve or climate data, see \cref{sec:methods-computation-p-out}. Furthermore, in \cref{sec:appendix-validation-poweroutput} we show all metrics derived when using observed time series for power output instead of simulated generation.

    Note that the factors in \cref{eq:multiplicative-decomposition}, are ratios of averages and not averages of ratios. A ratio of averages can be interpreted as weighted averages (see \cref{sec:appendix-ratio-of-avgs}). Output power density and input power density are average power output and input power, weighted by rotor swept area. System efficiency is weighted by power input of each turbine. This means that turbines with larger rotor swept area have a larger influence on the resulting time series, which is not the case for other measures of efficiency, such as the average coefficient of power (see \cref{sec:appendix-alternative-definitions-efficiency}).

\section{Results\label{sec:results}}
    \begin{figure}
        \centering{}
        \includegraphics[width=1\textwidth]{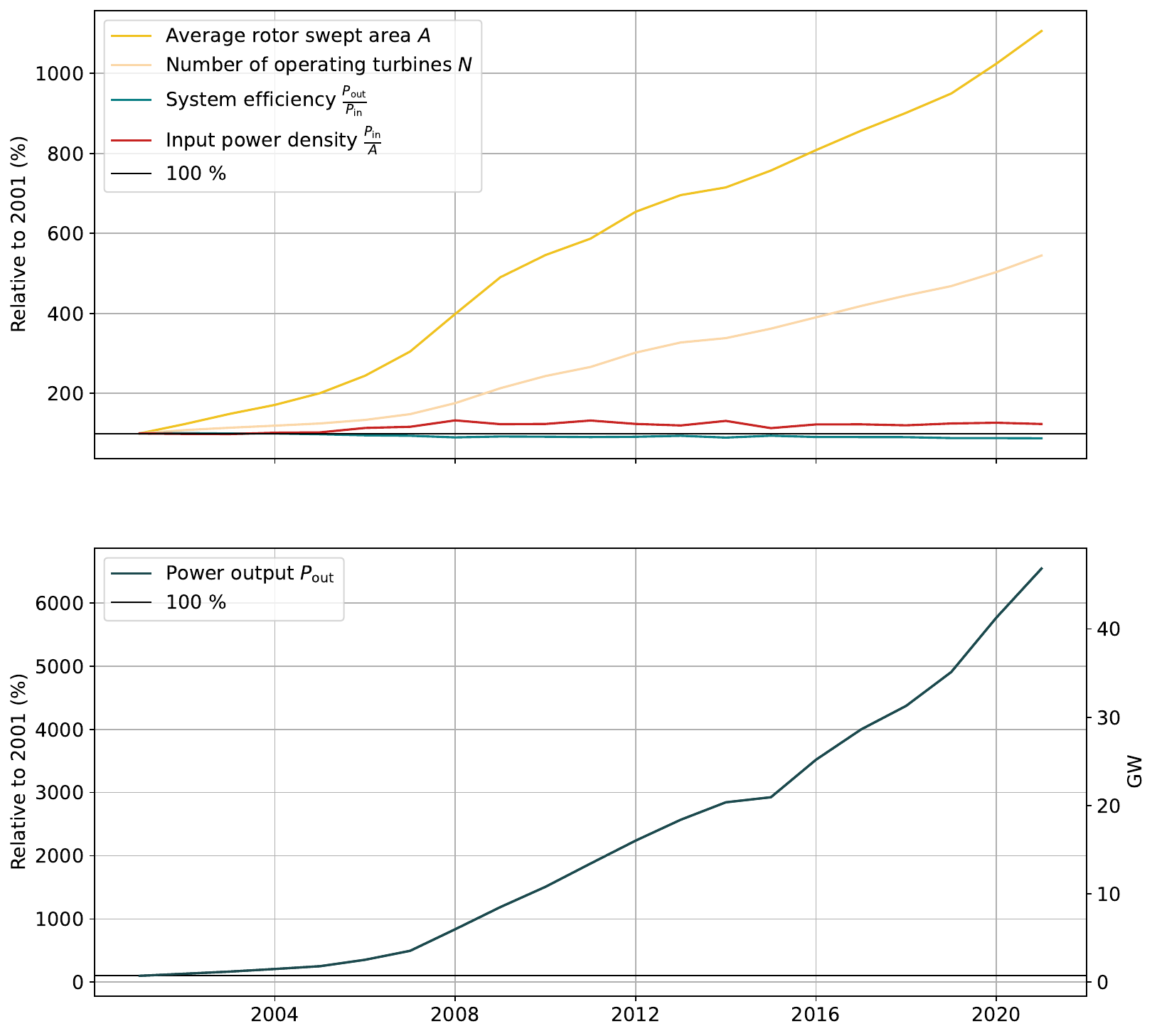}
        \caption{Driving factors of wind power generation: number and rotor size of turbines are the major drivers for the growth of wind power generation (system efficiency and power input power density are displayed in different scaling in \cref{fig:system_efficiency} and \cref{fig:p_in_with_trends}, respectively). The upper figure shows the factors on the right hand side of \cref{eq:multiplicative-decomposition}, the lower figure shows the left hand side of the equation.}
        \label{fig:p_out_decomposition}
    \end{figure}

    The growth of wind power generation during the last decades was mostly driven by an increasing number and size of turbines. If only the number of operating turbines had changed, all other factors in \cref{eq:multiplicative-decomposition} equal, wind power generation in the US would have increased by \qtyincludevalue{growth_num_turbines_built}{\%} since 2001. If only rotor swept area had changed, wind power output would have increased by \qtyincludevalue{growth_rotor_swept_area_avg}{\%} (see \cref{fig:p_out_decomposition}).
    The number of operating turbines\footnote{Decommissioning of turbines is neglected here, but it has a minor effect (see \cref{sec:methods-data-turbines}).} has increased from \input{./data/output/data-values/number-of-turbines-start}\unskip in 2001 to \input{./data/output/data-values/number-of-turbines-end}\unskip in 2021. The average rotor swept area of a turbine has increased from \qtyincludevalue{rotor_swept_area_avg-start}{m^2} in 2001 to \qtyincludevalue{rotor_swept_area_avg-end}{m^2} in 2021, caused by an increase in average blade length.

    Besides number and size of turbines, wind power output is affected by input power density and the system efficiency (see \cref{eq:multiplicative-decomposition}). Output power density is the combination of input power density and system efficiency as shown in \cref{eq:power-output-decomposition}. These three characteristics will be analyzed in the following subsections.

    Note that a validation of simulated power output with observed data suggests a greater confidence of results in the period after 2009, see \cref{sec:discussion} and \cref{sec:appendix-validation-poweroutput}.

\subsection{Output power density\label{sec:results-output-power-density}}

   \begin{figure}
        \centering{}
        \includegraphics[width=1\textwidth]{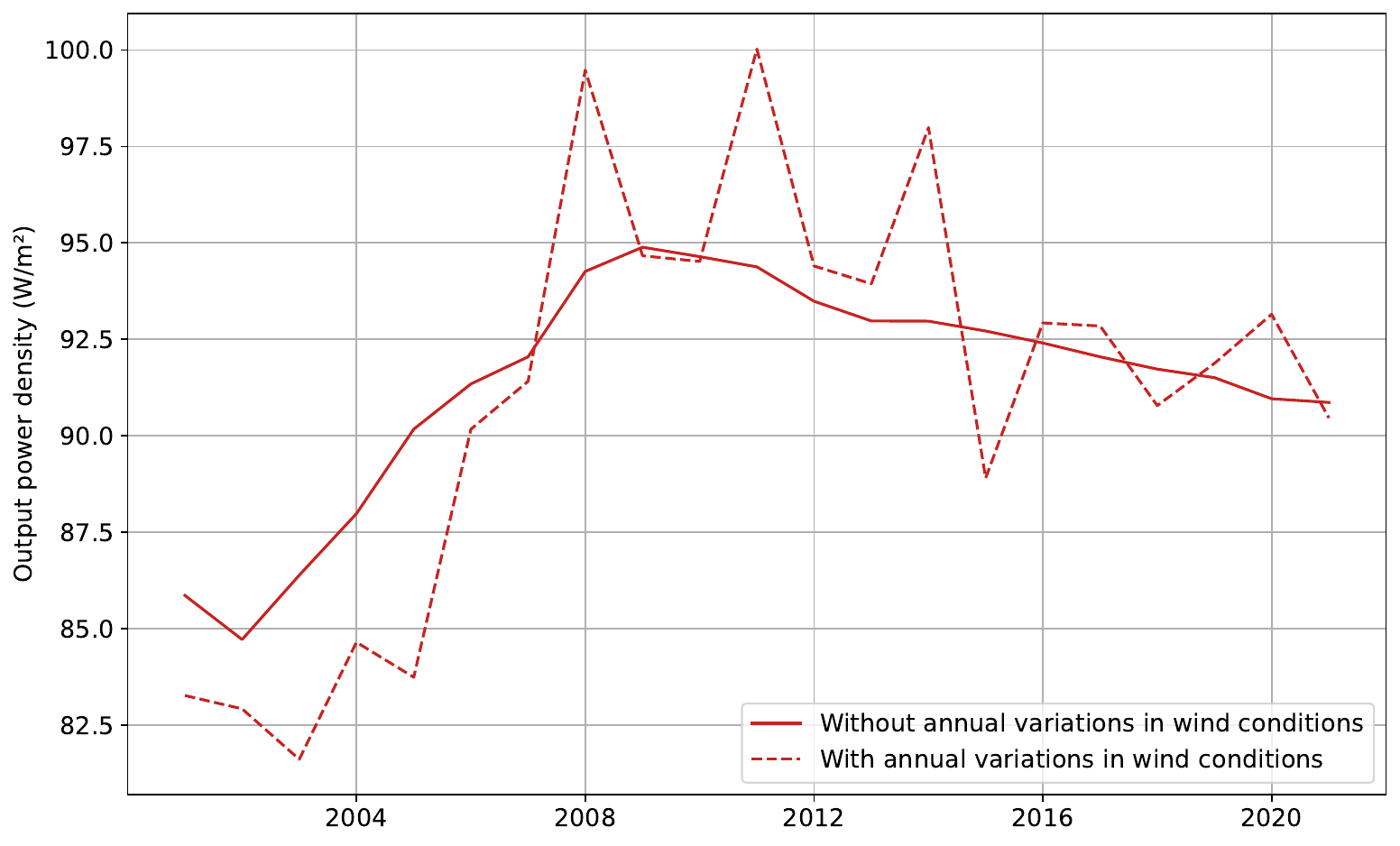}
        \caption{Output power density $\frac{\Pout}{A}$: electricity generation per unit of rotor swept area shows strong annual variations, an increase from 2001 to 2009 and a slight decline in the period since 2009.}
        \label{fig:d_out}
    \end{figure}

    As expected, generated electric power per rotor swept area, i.e. the output power density, is subject to strong annual variations. In years with high average wind speeds, wind power output is higher than in years with lower average wind speeds. Using long-term average wind power to calculate output power density allows identifying the underlying trends without variations caused by annual changes in wind conditions (see \cref{sec:methods-computation-p-out}). In this scenario, output power density increased from \qtyincludevalue{d_out_avgwind_start}{W/m^2} to \qtyincludevalue{d_out_avgwind_max}{W/m^2} in the period between 2001 and \input{./data/output/data-values/d_out_avgwind_idxmax}\unskip. Since \input{./data/output/data-values/d_out_avgwind_idxmax}\unskip output power density then dropped to \qtyincludevalue{d_out_avgwind_end}{W/m^2} in 2021 (\cref{fig:d_out}).
    This implies that in 2021, around \qtyincludevalue{d_out_less_percentages_end}{\%} less power output was generated per unit of rotor swept area than in \input{./data/output/data-values/d_out_avgwind_idxmax}\unskip. Annual variations in output power density due to changes in average wind speeds range from \qtyincludevalue{d_out_variations_min}{W/m^2} to \qtyincludevalue{d_out_variations_max}{W/m^2} when compared with the scenario using long-term average wind power.

    Output power density is a combination of two opposing trends (see \cref{eq:power-output-decomposition}). As shown in the next sections, system efficiency declines, but input power density shows an increase since 2001 and stabilizes after \input{./data/output/data-values/d_out_avgwind_idxmax}\unskip. In combination, additional input power density was able to offset the decrease of system efficiency until \input{./data/output/data-values/d_out_avgwind_idxmax}\unskip, but between \input{./data/output/data-values/d_out_avgwind_idxmax}\unskip and 2021 the decline of system efficiency prevails.

\subsection{System efficiency\label{sec:results-efficiency}}

    \begin{figure}
        \centering{}
        \includegraphics[width=1\textwidth]{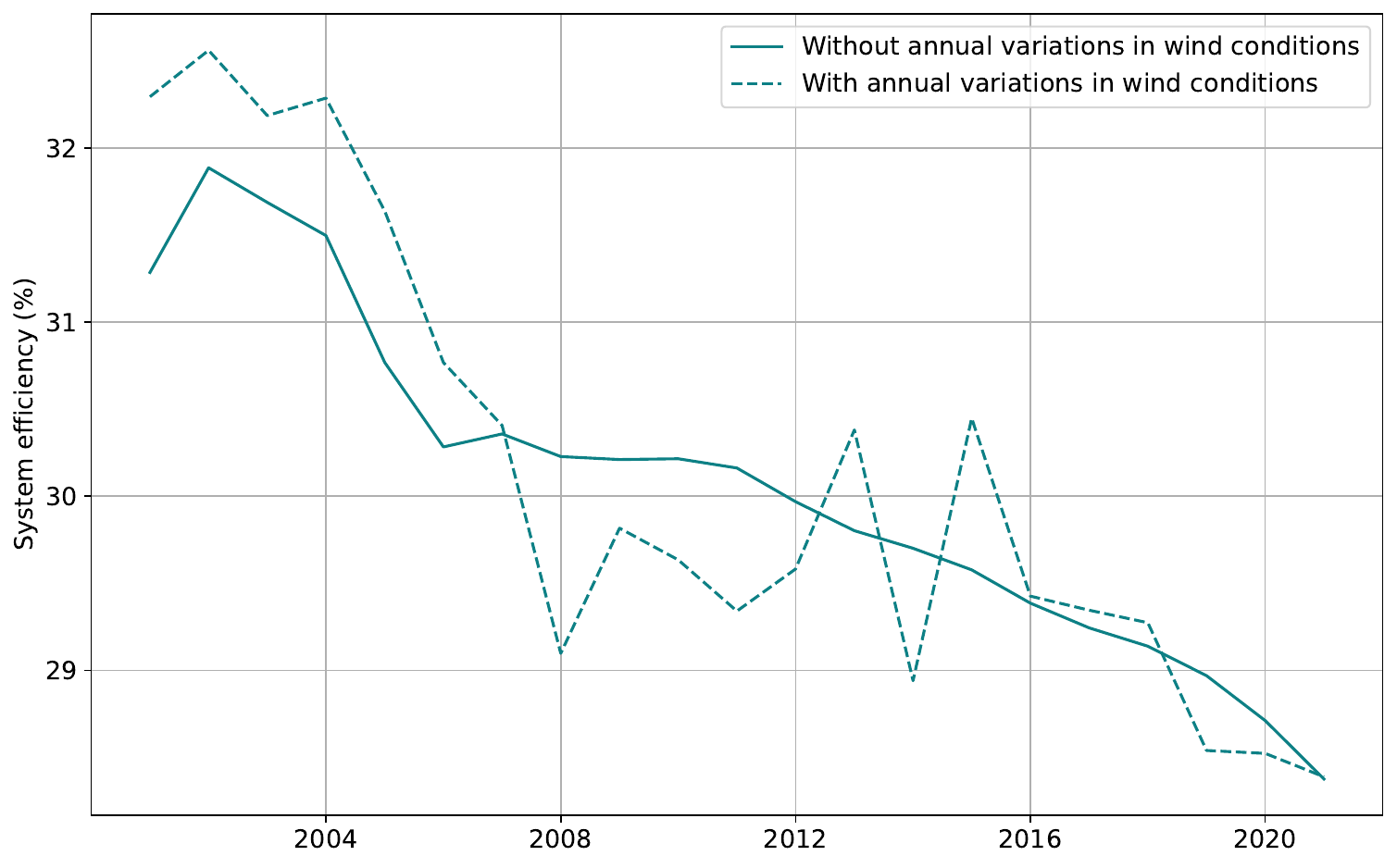}
        \caption{System efficiency $\frac{\Pout}{\Pin}$ shows a decline since 2001 with substantial variability between years due to fluctuations in wind conditions. Annual variations can be eliminated by calculating a theoretical scenario with long-term average wind power.}
        \label{fig:system_efficiency}
    \end{figure}

    System efficiency, i.e. the share of power output compared to power input, has declined over the past two decades from \qtyincludevalue{efficiency_start}{\%} in 2001 to \qtyincludevalue{efficiency_end}{\%} in 2021, with substantial variability between years (\cref{fig:system_efficiency}), caused by annually changing wind conditions. Annual variations are eliminated when system efficiency is calculated from power input and power output using long-term average power in the wind. In this scenario, system efficiency decreases \input{./data/output/data-values/efficiency_avgwind-decline-per-year}\unskip percentage points per year on average.

    System efficiency, as defined here, is subject to various characteristics of wind power generation. Firstly, system efficiency is negatively correlated to input power density. In times of low wind speeds, system efficiency tends to increase, while it tends to decrease in years with high wind speeds.
    The reason is that the system efficiency of converting power in the wind into electricity is not uniform over the whole range of wind speeds for wind turbines.
    For low wind speeds, this conversion efficiency is low, it increases with higher wind speeds and decreases again for very high wind speeds. For the US, we find that relatively low wind speeds move the fleet of wind turbines into a more optimal range of the conversion efficiency, while higher wind speeds tend to move the fleet out of that range. This effect is partly explained by temporal and spatial aggregation, as high power input or power output values have a larger impact on the average as defined here (see \cref{sec:appendix-ratio-of-avgs}).
    In conclusion, we find that the system efficiency is lower in wind resource-rich years. The correlation coefficient between yearly aggregated time series of system efficiency and input power density of all turbines is $\rho=\input{./data/output/data-values/correlation-efficiency-vs-input-power-density}\unskip$ (see also \cref{fig:scatter_efficiency_input_power_density} in the appendix).

    Besides input power density, system efficiency is largely affected by specific power. Lower specific power\footnote{Specific power is the ratio between the nameplate capacity, i.e. the size of the generator, and the rotor swept area of a wind turbine.} can reduce system efficiency, as it increases the share of the input wind power, which is not converted to electricity by wind turbines compared to a turbine with higher specific power.
    At the same rotor size, a turbine with a smaller rated capacity but similar slope in its power curve will reach the upper limit of the turbine capacity more often than one with a higher rated capacity -- this implies that implicitly, input wind power is curtailed by turbines with a lower specific power. We find that the specific power decreased from \qtyincludevalue{specific-power-start}{W/m^2} to \qtyincludevalue{specific-power-end}{W/m^2} during 2001--2021 (see \cref{fig:growth_and_specific_power}d), which is in accordance with other results \cite{bolinger_opportunities_2020}. Hence, we conclude that the decline of specific power is one of the reasons for lower system efficiency.

    Other factors can affect system efficiency too.
    However, lack of detailed data, and the high correlation between factors prevent us from drawing conclusions on their individual contribution to the change in overall system efficiency. A change in efficiency can be caused by aging effects of turbines, which are not included in our main model. However, we tested if the identified trends in our results changed when adding a aging loss correction factor (see \cref{sec:appendix-aging-effects}), and our analysis is robust to these changes.
    Wake effects, curtailment of wind energy, downtime due to maintenance, and downtime due to extreme weather conditions other than wind speeds such as icing or snow also reduce system efficiency. In our computation these effects are assumed to be constant over time and are controlled for by adding a constant loss factor. This assumption is discussed in detail in \cref{sec:discussion}, \cref{sec:methods-computation-p-out} and \cref{sec:appendix-validation-poweroutput}.

     All other losses can be summarized as technical efficiency, i.e. losses in the electrical generator and transmission or mechanical losses. These losses can be expected to decrease due to technical progress and therefore may partially offset the decline of system efficiency caused by specific power. The extent to which technological progress is correctly identified by our power curve model, is uncertain\footnote{Specific power is a parameter in the power curve model we use and due to the strong correlation between time and specific power, in average modern turbine models have a lower specific power. As the power curve model is estimated from existing power curves using a regression model, a lower specific power may implicitly imply a higher technical efficiency as mainly newer turbines will be used in the estimation. As this is implicit, we cannot understand to which extent technological progress is included in our computation of system efficiency.}. Nevertheless, a comparison with reported wind power output confirms our results (see \cref{sec:appendix-validation-poweroutput}). We conclude, that technological progress is either not significant for the declining trend of system efficiency or that it is properly reflected in the used power curve model.

\subsection{Input power density \label{sec:results-input-power-density}}

    \begin{figure}
            \centering{}
        \includegraphics[width=1\textwidth]{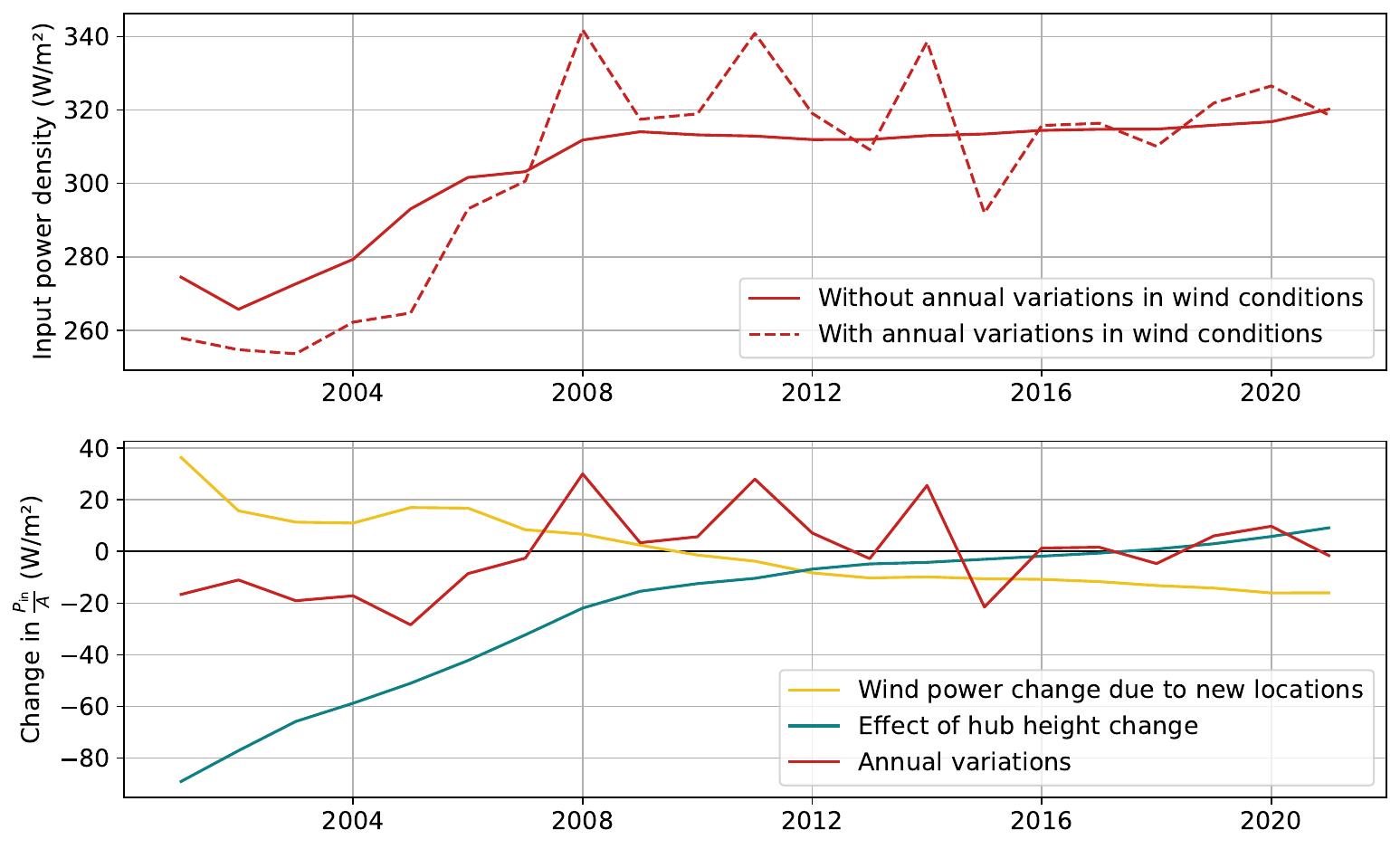}
        \caption{Input power density, i.e. how much wind power flows through rotor swept area of operating turbines normalized by area, could be increased since 2001. This is mostly caused by higher turbines and therefore higher wind speeds at hub height. On average, new locations are less windy compared at a constant reference height.}

        \label{fig:p_in_with_trends}
    \end{figure}

    Input power density $\frac{\Pin}{A}$, the kinetic power in wind per rotor swept area of all operating turbines, increased from \qtyincludevalue{d_in_avgwind_start}{W/m^2} in 2001 to to \qtyincludevalue{d_in_avgwind_end}{W/m^2} in 2021 (without annual variations). Until 2008 input power density increased \qtyincludevalue{d_in_avgwind_increase_until_2008}{W/m^2} per year. After 2009 the yearly increase was only \qtyincludevalue{d_in_avgwind_increase_from_2009}{W/m^2} on average (see \cref{fig:p_in_with_trends}a).

    This trend is the result of a combination of opposing effects: the effect of higher wind speeds due to larger turbine towers and the effect of new locations with lower wind speeds. In addition, input power density is affected by annual variations of wind speeds at the turbines locations. An additive decomposition was used to quantify these effects (see \cref{sec:methods-decomposition-d_in} for further details). \cref{fig:p_in_with_trends}b shows the decomposition.

    As expected, the growth of average hub heights has a positive effect on input power density. Since 2001 an increase of \qtyincludevalue{d_in_effect-of-hub-height-change}{W/m^2} can be attributed to higher wind turbines, i.e. input power density would have increased by that amount if only hub heights changed. At the same time the deployment of turbines at locations with less wind resources, compared at the same reference height, had a negative effect on input power density. Between 2001 and 2021, input power density would have decreased by \qtyincludevalue{d_in_abs-change-new_locations}{W/m^2} if only changes due to new locations were considered. This result is, however, sensitive to the data set used for wind speed bias correction: the trend is clear when using \qty{50}{m} and \qty{100}{m} wind speeds from the global wind atlas, but vanishes once the \qty{200}{m} version is used.
    Changes in input power density due to annual variations of wind conditions are of larger amplitude and range from \qtyincludevalue{d_in_variations_min}{W/m^2} to \qtyincludevalue{d_in_variations_max}{W/m^2}.

\section{Discussion\label{sec:discussion}}

The rotor swept area indicates the spacing area necessary in-between turbines, and thus the size of land necessary for deploying a wind park. If the distance between turbines required to reduce wake effects to an acceptable level is proportional to the rotor diameters, the total rotor swept area is also proportional to the spacing area if effects at the borders of wind parks are neglected. A simplified rule of thumb based on rotor sizes is often used to estimate spacing area requirements: it is assumed that a spacing of 7--12 times the rotor diameter in prevailing wind direction and 2--5 times the rotor diameter in the perpendicular direction between wind turbines has to be guaranteed \cites{ryberg_future_2018}{rinne_effects_2018}[76--77]{patel_wind_2006}[423]{manwell2010wind}. Lower distances will increase losses due to wake effects too much. Hence, output power density can be used as proxy for land use requirements for wind power generation.

 \citeauthor{miller_observation-based_2018} showed that power output per land area decreased from \qty{0.93}{W/m^2} in 2010 to \qty{0.90}{W/m^2} in 2016 \cite{miller_observation-based_2018,miller_corrigendum:_2019}. The decline in output power density identified by us is in accordance with their results. In \cref{fig:d_out_validation}, we show a direct comparison of \citeauthor{miller_observation-based_2018} to our results by assuming a fixed ratio between rotor diameter to the spacing between turbines in prevailing wind direction and perpendicular to it. In addition to confirming the results by  \citeauthor{miller_observation-based_2018} and extending their analysis for a longer time period, we furthermore deepen our understanding of the reasons for increasing land-use requirements: falling system efficiency, which outperforms increases in input power density in the last ten years, is identified as the main reason. The falling specific power of US wind turbines is one of the drivers of falling system efficiency.

Our approach of simulating wind power generation allows to simulate theoretical scenarios, such as wind power production using long-term average wind conditions, which enables us to remove noise due to annual variations in wind speeds. Nevertheless, this method does not allow to derive conclusions about losses due to curtailing, downtime, wake effects and aging of turbines.
We therefore introduced a constant loss factor to account for these effects which are not reflected otherwise in the estimation of power output via a power curve model (see \cref{sec:methods-computation-p-out}), while we show that the effects of aging do not have a strong effect on the trends observed in our results (\cref{sec:appendix-aging-effects}).

We furthermore validated the simulated time series for power output against measured power output time series provided by the EIA \cite{eia_wind_power_generation} and created a separate model to estimate output power density and system efficiency using the EIA data for power output, instead of our power curve simulations (see \cref{sec:appendix-validation-poweroutput}). Both approaches confirm a decrease in system efficiency and a stabilization or slight decrease in output power density in the period 2008--2019. However, the models deviate in the years before 2008 and after 2019. EIA data reflect earlier mentioned losses and technological progress, but the approach of simulating power output using power curves allows for more internal consistency, as errors in the US Wind turbine database are canceled\footnote{The simulated model uses the USWTDB to compute $A$, $\Pout$ and $\Pin$. On the other hand, the USWTDB is not used for $\Pout^\mathrm{EIA}$. If turbines are completely missing in the USWTDB, if decommissioned turbines are not correctly removed or if metadata is wrong, the error is smaller in $\nicefrac{\Pout}{\Pin}$ and $\nicefrac{\Pout}{A}$ than in $\nicefrac{\Pout^\mathrm{EIA}}{\Pin}$ and $\nicefrac{\Pout^\mathrm{EIA}}{A}$.}. In contrast, the observation approach captures all relevant effects, but missing turbines or missing or flawed metadata in the US Wind Turbine Database and errors in the observational wind power generation data may contribute to noisy results (see \cref{sec:appendix-validation-missing_data}).

We consider other limitations of minor importance, but list them here anyway. Turbine characteristics are missing for some turbines in the USWTDB. Data imputation techniques were applied to estimate missing values. Some turbines had to be removed completely during the data cleaning process, because the year of commissioning is missing or the turbine's specific power is not in a plausible value range.
Furthermore, to estimate input power, a constant value for air density has been used. It has been shown that variations in air density over space and time can lead to errors, when an average air density is used for simulating wind power output \cite{ulazia_global_2019}.
However, a validation of simulated power output using constant air density against observation data showed a good fit of simulation to observation, indicating that errors from the assumption of constant air density are low \cite{gruber2020global}.

\section{Conclusions\label{sec:conclusions}}

Concluding, we have shown that wind power was not able to increase its power output harmonized by rotor swept area -- this is in stark contrast to solar PV, where power generation per  panel area doubled in the past decade \cite{bolinger_pv_2022}. This implies higher specific land-use requirements and higher externalities linked to wind turbine rotors. We identify that higher land-use requirements are driven by declining system efficiency, which itself is partly a result of a fall in specific power. Increasing availability of input wind power was not sufficient to offset this decrease in system efficiency. At the moment, the market clearly favors wind turbines which are less efficient in using the wind available to them -- but also are less costly in terms of levelized costs of electricity (LCOE) and provide more benefits to the power system.
If specific power decreases further, windy locations become less available, and further increases in hub-heights are limited, input power density for new projects will stabilize or even decrease, and output power density will therefore also decrease further. This implies that the area swept by rotors required to produce one unit of electricity will increase, as will land-use requirements of onshore wind power - we therefore identify an inherent trade-off between minimizing cost and minimizing land-use of wind turbines.

\section{Methods \& Data\label{sec:methods}}

    In the following sections, details about the computation of the presented results and used data sets are described. Our code is published under the MIT license on Github\footnote{https://github.com/inwe-boku/windpower-decomposition-usa} and can be used to reproduce the results.

\subsection{Computation of power input\label{sec:methods-computation-p-in}}

    In a first step, wind velocities at \qty{10}{m} and \qty{100}{m} at the precise turbine location were estimated from raw ERA5 data (see \cref{sec:methods-data-wind-speeds}) using bilinear interpolation in the directions of longitude and latitude. Directionless wind speeds were then computed from u and v components for each turbine location $l$ and hour $t$.

    Subsequently, the wind power law $v_h = v_{h_0} \cdot \left(\frac{h}{h_0}\right)^\alpha$ was applied to estimate the wind speed at different heights $h$ by computing
    \[
        \alpha = \log_{\frac{\qty{100}{m}}{\qty{10}{m}}}\left({\frac{v_{\qty{100}{m}}}{v_{\qty{10}{m}}}}\right)
               = \frac{\log\left({\frac{v_{\qty{100}{m}}}{v_{\qty{10}{m}}}}\right)}{\log\left(10\right)}
    \]
    for each turbine location and time stamp and then calculate $v_h$ with $h_0 = \qty{100}{m}$.

    In the following we will use $v_{t,l}$ to denote the wind speed at time stamp $t$ and turbine location $l$. The height $h$ is omitted here in the notation for simplicity. The calculation of power input is done at a constant reference height and at hub heights of the installed turbines (see \cref{sec:methods-decomposition-d_in}).

    To increase spatial resolution, a bias correction was applied to the wind speeds using average wind speed data provided by the Global Wind Atlas 2 (GWA2). Bias correction factors are given by
    \[
        B_l = \frac{
            \left| T \right| \cdot v_{l,\qty{100}{m}}^{\mathrm{GWA2}}
        }{
            \sum_{t \in T} v_{t,l,\qty{100}{m}}^{\mathrm{ERA5}}
        }.
    \]
    Here, $T$ represents all hourly time stamps in the given period, $v_{t,l,\qty{100}{m}}^{\mathrm{ERA5}}$ is the uncorrected wind speed at location $l$ and time $t$ at 100 meter above surface calculated from the ERA5 data set and $v_{l,\qty{100}{m}}^{\mathrm{GWA2}}$ is the mean wind speed at location $l$ taken from the GWA2 data set.
    The corrected wind speed at time stamp $t$ and location $l$ is given by
    \[
        v_{t,l} = B_l \cdot v_{t,l}^{\mathrm{ERA5}},
    \]
    where $v_{t,l}^{\mathrm{ERA5}}$ is wind speed derived from ERA5 at hub height or reference height as explained above.

    To compute $\Pin$, we apply the formula $\pin \left(v_{t,l}\right) = \frac{1}{2} \rho A_l v_{t,l}^3$. $\Pin$ is the yearly aggregated time series for all turbine locations $L$
    \[
        \Pin\left(Y\right)
            = \frac{1}{\left|Y\right|} \sum_{t\in Y} \sum_{l\in L} b_{Y,l} \cdot \pin \left(v_{t,l}\right),
    \]
    where $Y$ is the set of hourly timestamps for each year (see also \cref{eq:p_in-formula}) and $b_{Y,l}$ indicates whether turbine $l$ was already operating in year $Y$. $b_{Y,l}$ is set to $0$ before the turbine was built and to $1$ when it is operating. In the commissioning year, we set $b_{Y,l} = 0.5$. This is motivated by the assumption that turbines are built with equal probability throughout the year, so on average, they contribute only 50\% to the aggregated value in the first year.

    Time series assuming long-term wind conditions are calculated by averaging over the whole time period of 2001--2021:
    \[
        P_{\mathrm{in,avg}}\left(Y\right)
            =  \sum_{l\in L} b_{Y,l} \frac{1}{\left|T\right|} \sum_{t\in T} \pin \left(v_{t,l}\right).
    \]
    That means the differences between years in $P_{\mathrm{in,avg}}\left(Y\right)$ arise only due to new turbines being built.

    The average power input at reference height of \qty{80}{m} $P_{\mathrm{in,refh,avg}}$ is calculated analogously by using wind speeds $v_{t,l}$ at reference height instead of hub height.

\subsection{Computation of power output \label{sec:methods-computation-p-out}}

    Analogous to the computation of $\Pin$, computation of power output is based on bias corrected wind speeds using ERA5 and GWA2 data. Here, wind speeds are estimated only at turbine hub heights. $\Pout$ is then calculated using the power curve model $\pout \left(v_{t,l}\right)$ by \citeauthor{ryberg_future_2018} (\cref{sec:methods-data-powercurve-model}). The power curve model takes specific power as input parameter and provides capacity factors for wind speeds. Specific power is calculated for each turbine $l$ using data provided by the USWTDB. Hence, we obtain a power curve for each turbine location which is used to calculate power output from wind speeds for each hour $t$ at location $l$. The aggregated time series for power output and power output under long-term average wind conditions is calculated analogous to power input as described in the previous section, but we add a constant loss factor $R$ to account for downtime and losses not reflected in the power curve model:

    \begin{align*}
        \Pout\left(Y\right)
            &= R \cdot \frac{1}{\left|Y\right|} \sum_{t\in Y} \sum_{l\in L} b_{Y,l} \cdot \pout \left(v_{t,l}\right)\\
        P_{\mathrm{out,avg}}\left(Y\right)
            &= R \cdot \sum_{l\in L} b_{Y,l} \frac{1}{\left|T\right|} \sum_{t\in T} \pout \left(v_{t,l}\right).
    \end{align*}

    We chose $R=\input{./data/output/data-values/loss_correction_factor}\unskip\%$ which is the least square fit between $\Pout$ and observed time series by the EIA in the period 2008--2021 both normalized by the total rotor swept area $A$ (earlier years are assumed to have a higher uncertainty, see \cref{sec:appendix-validation}). This is in accordance with the downtime factor of $85\%$ used in a simulation by \citeauthor{HOLTINGER201649} \cite{HOLTINGER201649} -- derived from several other sources.

\subsection{Computation of total rotor swept area and number of operating turbines\label{sec:methods-computation-turbines}}

    Time series for the total rotor swept area $A$, and number $N$ of operating turbines are derived from the United States Wind Turbine Database (USWTDB). After removing all turbines from the data set, where the commissioning year is missing, mean data imputation for each commissioning year was used to estimate missing values for the rotor diameter parameter (see \cref{sec:methods-data-turbines}). The rotor diameter was then used to calculate the rotor swept area $A_l$ for each turbine at location $l$. To compute a time series, the sum of the rotor swept areas of all operating turbines was calculated for each year.

    The commissioning date is given only with a yearly resolution. Therefore a weight of $0.5$ is used for turbines in their commissioning year when summing up rotor swept areas of operating turbines and the number of turbines in an analogous way as $b_{Y,l}$ is used to compute power input in \cref{sec:methods-computation-p-in}.

\subsection{Decomposition of input power density\label{sec:methods-decomposition-d_in}}

    Input power density $\frac{\Pin}{A}$ is subject to large climatic variations. An additive decomposition is used to observe underlying trends unrelated to the climate. In the following, we describe the decomposition in more detail.

    Note that input power density $\frac{\Pin}{A}$ is proportional to the weighted average of $v_{t,l}^3$ with weights $A_l$ (see \cref{sec:methods-computation-p-in}, note that a constant value is used for the air density $\rho$). Therefore, input power density changes precisely if wind speeds $v_{t,l}$ at locations with operating turbines change. Hence, changes in input power density are thoroughly explained, if changes can be attributed to different reasons for changing wind speeds at turbine locations.

    The average wind speed at the hub height at turbine locations changes because of turbines being added at new locations, a change of the average hub height, or a change of climate conditions (either annual or multi-annual variability or trends due to, e.g. climate change). In order to analyze these different components, power input is computed under different hypothetical conditions. $P_{\mathrm{in,refh,avg}}$ is the power input at turbine locations at a reference height of \qty{80}{m} using average power input over the entire period. $P_{\mathrm{in,avg}}$ is the total power input at all turbine locations at the hub height of installed turbines when also assuming average climate conditions. $\Pin$ is the actual power input captured by all installed wind turbines (see \cref{sec:methods-computation-p-in}). These definitions can be used for the following decomposition:
    \begin{align*}
    \Pin= &  & \textrm{baseline} & \qquad \nonumber \\
     & + & P_{\mathrm{in,refh,avg}} - \textrm{baseline} & \qquad\leftarrow\textrm{\scriptsize effect of new locations}\nonumber \\
     & + & P_{\mathrm{in,avg}}-P_{\mathrm{in,refh,avg}} & \qquad\leftarrow\textrm{\scriptsize hub height change}\label{eq:additive-decomposition}\\
     & + & \Pin-P_{\textrm{in,avg}} & \qquad\leftarrow\textrm{\scriptsize annual variations of climate conditions}\nonumber
    \end{align*}

    The baseline is an arbitrary value. It is chosen as the average of $P_{\mathrm{in,refh,avg}}$ over the complete time span. Similarly, the reference height of \qty{80}{m} is the median of all turbine hub heights. Both values, baseline and reference height, do not change the trends of the time series, but only add an offset. To get an additive decomposition of the input power density $\frac{\Pin}{A}$, we divide the equation on both sides by $A$. This allows quantifying the effect of changes in wind speeds due to new locations, the effect of change in hub heights and the effect of annual variations in available wind resources.

    Note that the annual variations of climate conditions is not independent of the effect of hub height change. Average wind speeds at hub height are increasing over time, since average hub heights of new turbines are increasing over time and average wind speeds are higher at larger heights. Therefore also the difference $\Pin-P_\textrm{in,avg}$ exhibits larger variations over time, when compared with variations of input wind power at the same heights. However, the due to the increasing number of operating turbines, annual variations at hub height decrease over time.

    A graphical explanation of the additive decomposition of input power density is given in the Appendix in \cref{fig:decomposition_pin-waterfall}.

\subsection{Data\label{sec:methods-data}}

    We aimed at mainly using publicly available data sets to compute the presented results. However, the turbine data set was extended by non-public data as explained in \cref{sec:methods-data-turbines} in more detail. External input data were validated using additional data sets as described in the Appendix in \cref{sec:appendix-validation}.

\subsubsection{Wind speed data\label{sec:methods-data-wind-speeds}}

    To estimate the power input at turbine locations, wind speeds from the ERA5 data set \cite{reanalysis_era5_single_levels} were used. ERA5 is an openly available global reanalysis data set. Wind velocities are provided at \qty{10}{m} and \qty{100}{m} height with hourly temporal resolution and 0.25° spatial resolution -- in the US this results in tiles with a size of approximately $\qty{25}{km} \times \qty{25}{km}$.

    To increase spatial resolution, the Global Wind Atlas Version 2 (GWA2) was used \cite{GWA2}. It provides average wind speeds at a resolution of 0.0025°, so each ERA5 tile is covered by $10\,000$ GWA2 tiles.

\subsubsection{Wind turbine data\label{sec:methods-data-turbines}}

    The United States Wind Turbine Database \cite{uswtdb} is a collection of \input{./data/output/data-values/number-of-turbines-end}\unskip wind turbines located in the USA. Every turbine is annotated with a precise location, i.e. longitude and latitude, and several other meta parameters such as hub height, rotor diameter, capacity, model name and commissioning year.

    However, many of the meta parameters are missing. For \input{./data/output/data-values/missing_ratio_rd_hh}\unskip\% of the turbines in the data set, there is no hub height or no rotor diameter available. Simply discarding turbines with missing data would lead to a significant underestimation of aggregated values such as the total rotor swept area of operating turbines, which is essential in most parts of the computation.
    Instead, we used mean data imputation for each year to estimate values for the missing meta parameters for hub height, rotor diameter and capacity. The parameters are missing not at random (MNAR), as there are more missing parameters for older turbines (see \cref{sec:appendix-validation-missing_data}).
    The distribution of missing parameters for turbines built in the same year is unknown, which is why there is no way to find an optimal way of estimating missing values.  However, computation of minimum and maximum introduced error -- similar to a sensitivity analysis -- shows that the introduced error can be neglected.

    The public USWTDB data set contains currently operating wind turbines only. This study aims to analyze the historical development of turbines, which naturally requires also knowledge about decommissioned turbines. Via personal communication, we received an extension to the USWTDB data set \cite{uswtdb_decommissioned_extension}, which consists of turbines that have been removed from the USWTDB by now due to their decommissioning. These turbines have been merged with the publicly available data set. However, there are turbines appearing in older versions of the USWTDB which are neither part of the latest version nor of the extension data set, which contains decommissioned turbines. We therefore merged the USWTDB versions 3.01, 4.1, 5.0 and 5.1 and then removed duplicates by using the longitude and latitude of the turbines. %

    For some of the turbines in the data set, the commissioning year is missing. These turbines cannot be used in the calculation of the time series and are therefore removed in a preprocessing step. This affects \input{./data/output/data-values/percentage_missing_commissioning_year}\unskip\% of all turbines or \input{./data/output/data-values/percentage_missing_commissioning_year_start}\unskip\% of turbines operating in 2001. We assume that these are mostly older and therefore smaller turbines, since the share of turbines other meta parameters missing is higher for the ones with commissioning year before 2008 (see \cref{sec:appendix-validation-missing_data}). Therefore, the discarded total capacity can be assumed to be below \input{./data/output/data-values/percentage_missing_commissioning_year}\unskip\% of the total capacity or below \input{./data/output/data-values/percentage_missing_commissioning_year_start}\unskip\% relative to capacity installed in 2001.

    Decommissioning dates are missing for most turbines. For only \input{./data/output/data-values/num_available_decommissioning_year}\unskip turbines a decommissioning year is available, but \input{./data/output/data-values/num_decommissioned_turbines}\unskip turbines are marked as decommissioned and further \input{./data/output/data-values/num_further_old_turbines}\unskip turbines were older than 25 years in 2021. However, since this affects mostly smaller turbines, the effect on the total installed capacity is quite small (see \cref{sec:appendix-validation} and \cref{fig:irena_capacity_validation}). Therefore, all turbines were used without taking decommissioning into account.

    One turbine with a capacity of \qty{275}{kW} located on the Mariana Islands has been removed from the data set because it significantly reduces the size of the bounding box of turbine locations and, therefore, also the size of required wind speed data.

\subsubsection{Power curve model\label{sec:methods-data-powercurve-model}}

A power curve maps wind speeds to the expected power output of a wind turbine. Here we used the power curve model introduced by \citeauthor{ryberg_future_2018} \cite{ryberg_future_2018}. They used power curves provided by turbine manufacturers to fit a model with specific power as parameter.
A lower specific power shifts the peak in system efficiency to lower wind speeds, i.e. a turbine with a low specific power can generate more electric power per unit of power input for lower wind speeds and a turbine with higher specific power can generate more electric power per unit of power input for higher wind speeds. %

The power curve model is provided via a table of two parameters \cite[Table A.6]{ryberg_future_2018}. Discrete steps and linear interpolation were used for specific power and wind speed to calculate the power output for a certain turbine and wind speed.

\bigskip

\emph{Acknowledgments.} We gratefully acknowledge support from the
European Research Council (reFUEL ERC-2017-STG 758149). Furthermore we would like to thank Joseph Rand and Ben Hoen for providing data on decommissioned turbines.

\newpage

\section*{Symbols}

\begin{lyxlist}{00.00.0000}
\item [$\rho$] air density, constant value of  \qty{1.225}{kg/m³} is used
    \item [$R$] loss factor to account for wake effects, downtime and other losses
    \item [$\Pin$] power input of all turbines %
        (at hub height)%
    \item [$P_{\mathrm{in,refh,avg}}$] Wind power at a reference height assuming long-term average power
    \item [$P_{\mathrm{in,avg}}$] Wind power at hub heights assuming long-term average wind conditions
    \item [$\Pout$] simulated wind power output of all turbines
    \item [$P_{\mathrm{out,avg}}$] simulated power output of all turbines assuming long-term average wind conditions
    \item [$\Pout^\mathrm{EIA}$] observed wind power output as reported by the EIA
    \item [$N$] number of operating turbines %
    \item [$A_l$] rotor swept area of a single turbine at location $l$, i.e. $A_l = \pi \cdot \frac{d_l^2}{4}$ for a turbine with rotor diameter $d_l$
    \item [$A$] total rotor swept area of all turbines $L$, i.e.$A=\sum_{l\in L}A_{l}$
    \item [$L$] set of all turbine locations
    \item [$l$] location of a single wind turbine
    \item [$v$] wind speed
    \item [$v_l^{\mathrm{GWA2}}$] mean wind speed at location $l$ in the Global Wind Atlas 2 (GWA2)
    \item [$v_{t,l}^{\mathrm{ERA5}}$] mean wind speed at location $l$ at time $t$ in ERA5
    \item [$v_{t,l}$] wind speed at location $l$ at time $t$
    \item [$C_\mathrm{p}\left(v\right)$] coefficient of power, ratio of power output and power input at wind speed $v$ for a specific turbine
    \item [$Y$] set of time stamps $t$ in a year
    \item [$\pin \left(v\right)$] power input at wind speed $v$ for turbine at location $l$ with rotor swept area $A_l$, i.e. $\pin \left( v \right) = \frac{1}{2}\rho A_l v^3 $
    \item [$\pout \left(v\right)$] power output at wind speed $v$ for turbine at location $l$ with rotor swept area $A_l$, i.e. $\pout \left(v\right)=\frac{1}{2}\rho A_l v^3 C_\mathrm{p}\left(v\right)$

\end{lyxlist}

\nocite{miller_observation-based_2018,miller_corrigendum:_2019}
\nocite{le_gourieres_chapter_1982}
\nocite{irena_capacity}
\nocite{wiser_wind_market_report2019}
\nocite{irena_renewable_costs}
\nocite{ziegler_lifetime_2018}
\nocite{irena_wind_power_output}
\nocite{ulazia_global_2019}

\printbibliography

%% file: appendix.tex
%auto-ignore
\appendix
\renewcommand\thefigure{\thesection.\arabic{figure}}
\setcounter{figure}{0}
\renewcommand\thetable{\thesection.\arabic{table}}
\setcounter{table}{0}
\renewcommand{\theequation}{\thesection.\arabic{equation}}
\setcounter{equation}{0}

\section{Supplementary information}

\subsection{Motivation of multiplicative decomposition of power output\label{sec:appendix-motivation}}

    Our decomposition

    \[
        \Pout =
            N \cdot
            \frac{A}{N} \cdot
            \frac{\Pin}{A} \cdot
            \frac{\Pout}{\Pin}
    \]
    is motivated by the generic formula describing the relationship between wind available to turbines and wind power generation:
    \[
        \pout\left(v\right) =\frac{1}{2} \rho A_l  v^3 C_\mathrm{p}\left(v\right).
    \]
    The wind power output of a single wind turbine $\pout\left(v\right)$ is proportional to the rotor swept area $A_l$, to the cube of the wind speed $v$ and to the air density $\rho$. Here $A_l$ denotes the rotor swept area of a single wind turbine at location $l$ and should not be confused with the total rotor swept area $A$ of all operating turbines. The share of wind power converted to electricity is typically denoted by $C_\mathrm{p}\left(v\right)$ and called the coefficient of power (see, e.g. \cite{le_gourieres_chapter_1982}).
    $C_\mathrm{p}\left(v\right)$ describes how efficiently a turbine can convert wind power to electricity at wind speed $v$. Input wind power at wind speed $v$ is given by $\pin\left(v\right) =\frac{1}{2} \rho A_l  v^3$, so that the coefficient of power can be written as the ratio of power output to power input
    \[
        C_\mathrm{p}\left(v\right) =
            \frac{\pout\left(v\right)}{\pin\left(v\right)}.
    \]
    An upper bound to $C_\mathrm{p}\left(v\right)$ is given by Betz' limit. For a single turbine, the function $\pout(v)$ is usually called power curve, where $A_l$ depends on the rotor diameter of the turbine model, $v$ is the wind speed at hub height and $C_\mathrm{p}\left(v\right)$ combines all other turbine specific characteristics. Air density $\rho$ depends on climate conditions, weather and locations \cite{ulazia_global_2019}, but in practice, a constant value is typically used for power curves and also in our analysis. Here we use $\rho  = 1.225\si{kg/m^3}$. That means that rotor swept area $A_l$, the cube of wind speed $v^3$ and efficiency $C_\mathrm{p}\left(v\right)$ are the only non-constant parts in $\pout \left(v\right)$. Also note that $\nicefrac{\pin\left(v\right)}{A_l}$ is proportional to $v^3$. This motivates the decomposition of power output into total rotor swept area, system efficiency and input power density as shown in  \cref{eq:multiplicative-decomposition}. Total rotor swept area is further decomposed into the number of operating turbines $N$ and the average rotor swept area per turbine $\frac{A}{N}$.

    Our definition of system efficiency $\frac{\Pout}{\Pin}$ includes curtailing, wake effects and other losses. This is in contrast to the coefficient of power whereas $C_\mathrm{p}\left(v\right)$ takes only mechanical, electrical or aerodynamic losses caused by a single turbine into account (see \cref{sec:results-efficiency} for more details). However, system efficiency is the relevant measure to analyze the change in the total growth of wind power generation and its impacts.

\subsection{Validation\label{sec:appendix-validation}}

    We compared the data used in our analysis to different, independent sources in order to gain confidence in its validity. The following subsections describe deviations between data sources and related results by other authors and explain why we are confident that our conclusions are valid even under conservative assumptions despite the fact that some results disagree.

\begin{figure}[h]
        \centering{}
        \includegraphics[width=1\textwidth]{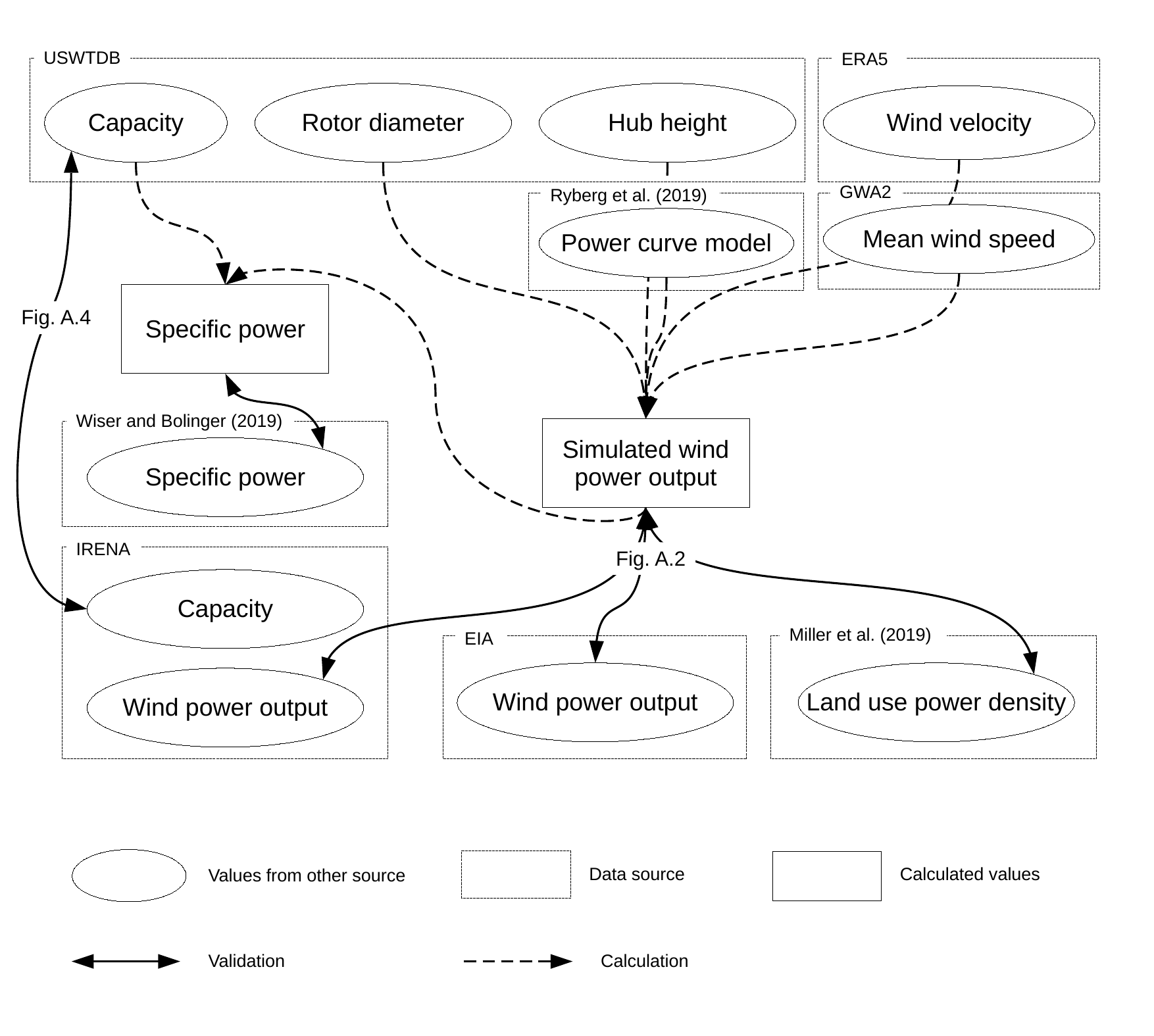}
        \caption{Overview of validation between different data
        sources.\label{fig:overview-validation}}
    \end{figure}

    \cref{fig:overview-validation} provides an overview of comparisons. Values from eight different sources are compared. A validation of our simulated power output time series is discussed in \cref{sec:appendix-validation-poweroutput}. The completeness of turbines in the USWTDB is analyzed in \cref{sec:appendix-validation-uswtdb-irena}. Furthermore, specific power calculated from the USWTDB is in accordance with the data reported in \cite{wiser_wind_market_report2019} (note that \cref{fig:growth_and_specific_power} shows specific power of all operating turbines, while \cite{wiser_wind_market_report2019} calculates the specific power of all new installed turbines).

\subsubsection{Validation of simulated power output\label{sec:appendix-validation-poweroutput}}

    Our results for simulated wind power output is validated with observed power output data by the EIA \cite{eia_wind_power_generation} and IRENA \cite{irena_wind_power_output}. A similar simulation of US wind power output by \citeauthor{gruber2020global} \cite{gruber2020global} already showed that bias corrected ERA5 wind speed data is able to represent wind conditions at wind turbine sites well, when estimating power output using the USWTDB and the power curve model by \citeauthor{ryberg_future_2018}. In \cite{gruber2020global} simulated power output is compared to monthly power output reported by the EIA. The RSME between monthly capacity factors in the period 2010--2019 was found to be 0.031.

\begin{figure}[h]
        \centering{}
        \includegraphics[width=1\textwidth]{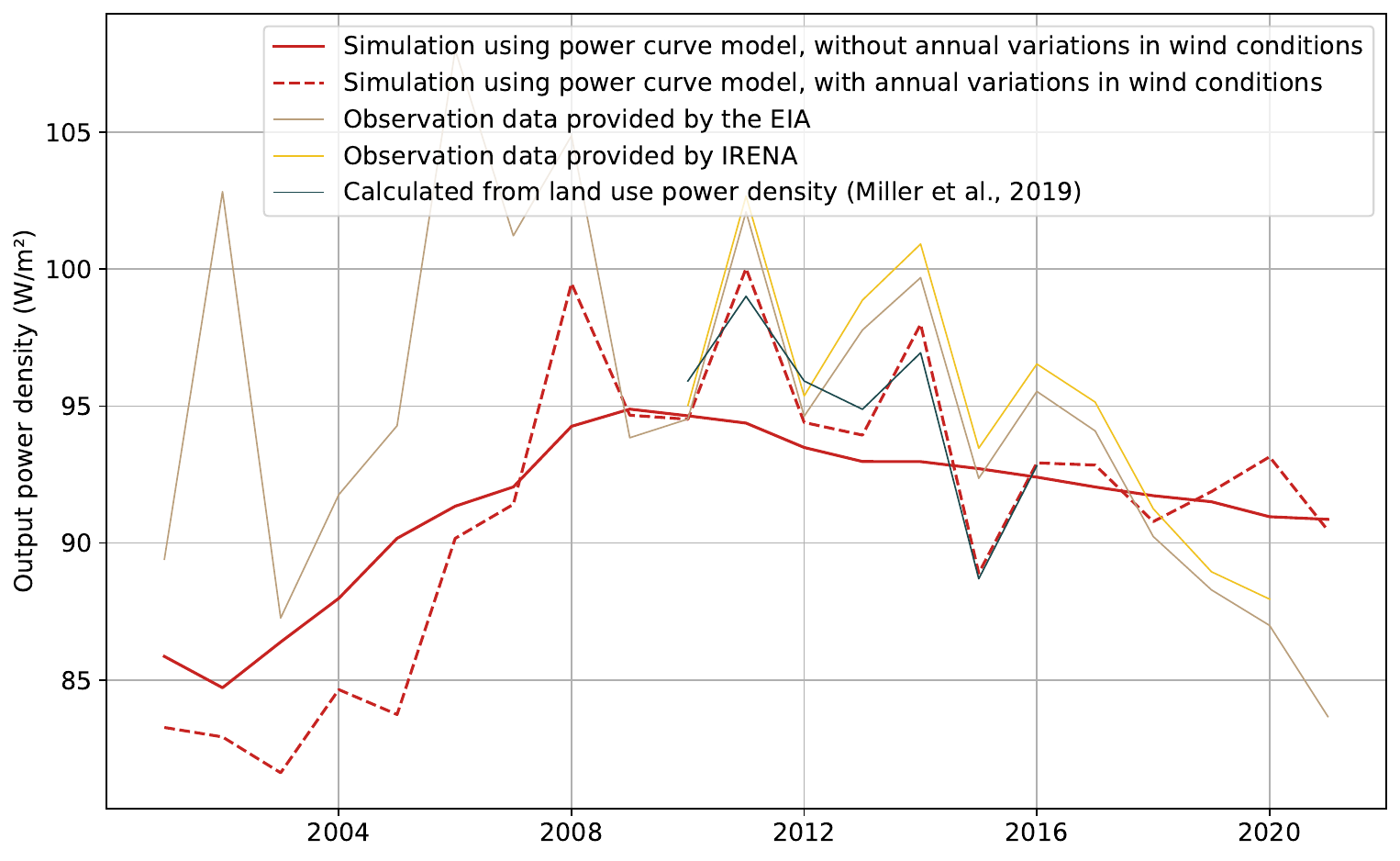}
        \caption{Comparison of output power density calculated first, using different time series for power output and rotor swept area from the USWTDB and secondly, by converting landuse power density results assuming a turbine spacing of 6Dx13.5D.\label{fig:d_out_validation}}
    \end{figure}

\begin{figure}[h]
        \centering{}
        \includegraphics[width=1\textwidth]{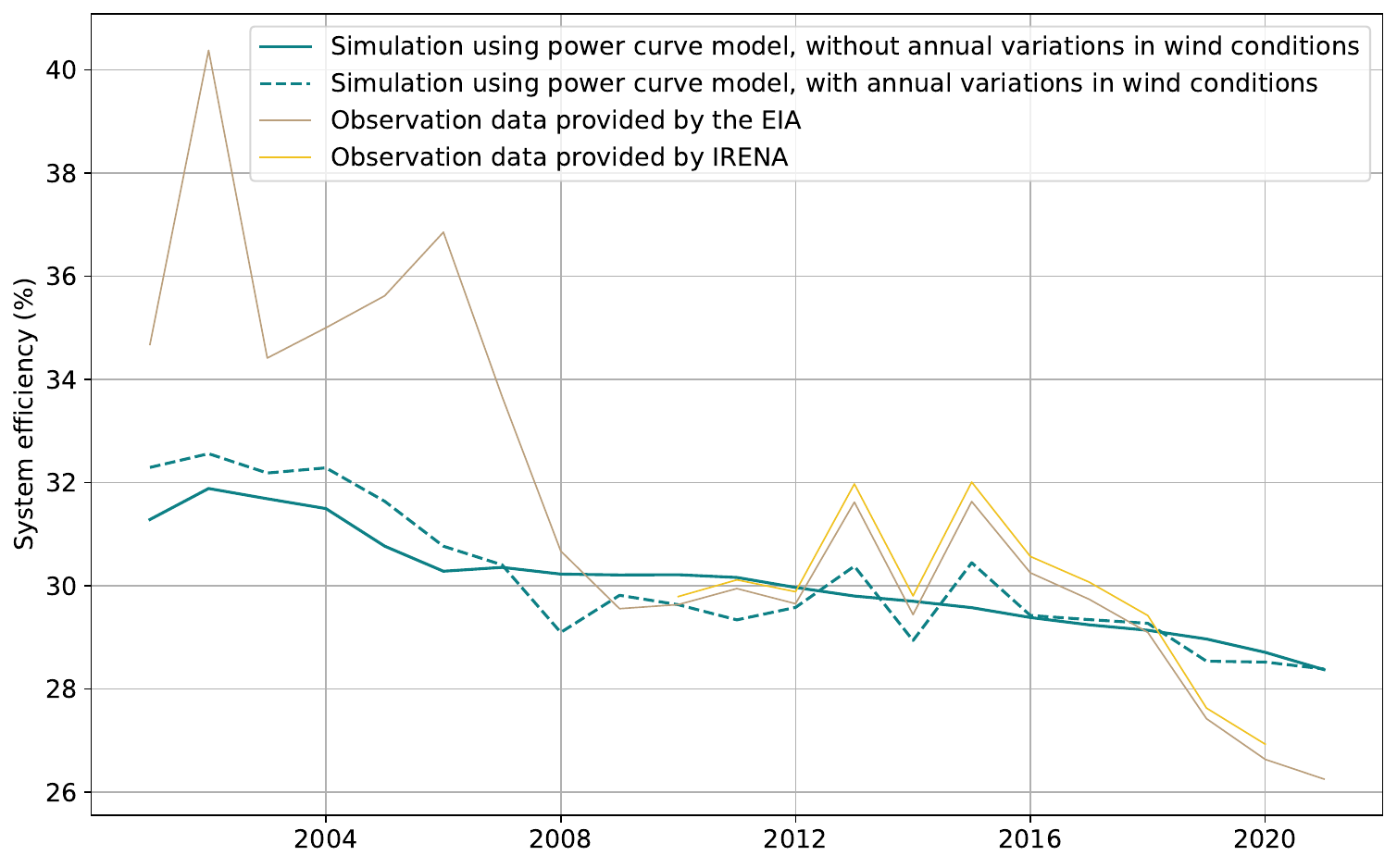}
        \caption{System efficiency calculated using different data for power output.\label{fig:efficiency_validation}}
    \end{figure}

    \Cref{fig:d_out_validation} and \cref{fig:efficiency_validation} compare results for output power density and system efficiency using different data sources for power output. In the years before 2008 and after 2019 there is a larger deviation between results using simulated power output and observed power output as reported by the EIA. The identified trends in system efficiency and output power density using simulated power output are less pronounced in the results using observed values. Data on power per land use by \citeauthor{miller_corrigendum:_2019} \cite{miller_corrigendum:_2019} have been converted to power per unit of rotor swept area assuming a spacing of 6Dx13.5D between turbines, where $D$ is the rotor diameter. This is slightly above the range of values of distance factors found in literature (see \cref{sec:discussion}). IRENA shows only very small deviations from EIA data, hence, we focus on a comparison between EIA data and our simulation in the following, because IRENA provides data only in the time range between 2010--2020.

    The deviation between simulated and observed power output can be caused by multiple effects, which will be listed in the following. Firstly, we assume that there is no significant bias in the wind speed data or that EIA data is flawed before 2008 or after 2019. A possible error could be introduced by our assumption that losses are constant over time (see \cref{sec:methods-computation-p-out}). If this is the case, EIA data suggests that losses due to wakes, curtailing and downtime were lower than average in the years before 2008 and larger than average in the years after 2019. Another possible cause for deviations lies in the power curve model. The power curve model could be biased in a way that it underestimates power output for old turbines, which does not explain the high variance of the EIA data. Lastly, turbine data from the USWTDB could be incomplete or flawed. If mostly old turbines are missing, it would explain the overestimation of $\nicefrac{\Pout^\mathrm{EIA}}{A}$. However, this would also imply that the IRENA capacity database is incomplete in the years before 2005 (see \cref{sec:appendix-validation-uswtdb-irena}). A bias in missing meta data could distort the data imputation used to infer missing parameters of turbines. But a bias which leads to an underestimation of total rotor swept area means that mostly large turbines are lacking characteristics in the USWTDB, which is not very likely the case.

    All possible errors which are related to the USWTDB or wind speed data can be assumed to be smaller in $\nicefrac{\Pout}{A}$ and in $\nicefrac{\Pout}{\Pin}$ than in $\nicefrac{\Pout^\mathrm{EIA}}{A}$ and $\nicefrac{\Pout^\mathrm{EIA}}{\Pin}$, because errors cancel out partially. We therefore assume that simulated results of output power density and system efficiency closer reproduce real trends. However, there is a greater confidence in results in the years after 2008 and before 2019.

\subsubsection{Validation of USWTDB completeness with total installed capacity reported by IRENA\label{sec:appendix-validation-uswtdb-irena}}

    Missing turbines in the USWTDB would lead to an underestimation of power input $\Pin$ and total rotor swept area $A$ and subsequently to an overestimation of output power density $\nicefrac{\Pout^\mathrm{EIA}}{A}$ and system efficiency $\nicefrac{\Pout^\mathrm{EIA}}{\Pin}$, when calculated using EIA data for power output. This would explain the deviation of results using simulated power output and observed power output in \cref{fig:d_out_validation} and \cref{fig:efficiency_validation}.

\begin{figure}[h]
        \centering{}
        \includegraphics[width=1\textwidth]{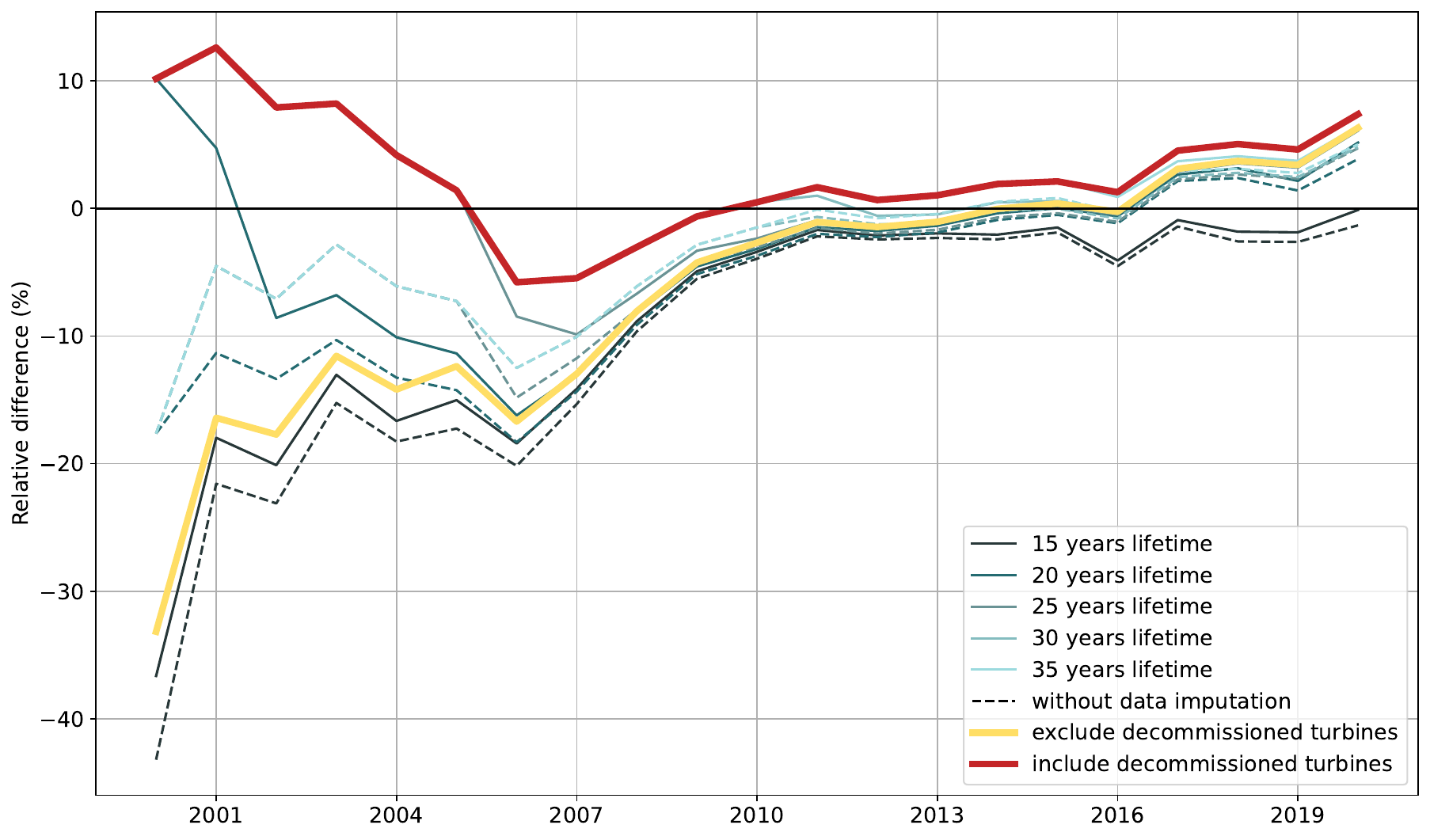}
        \caption{Comparison of installed wind power capacity data as provided by the USWTDB and IRENA. A positive value indicates that USWTDB reports a higher value for installed wind power capacity.}
        \label{fig:irena_capacity_validation}
    \end{figure}

    Therefore, we compared the USWTDB to a second data source. The  IRENA Query Tool \cite{irena_capacity} provides the yearly installed wind power capacity per country and technology. \cref{fig:irena_capacity_validation} shows the relative difference of installed wind power capacity between the two data sources IRENA and USWTDB. Installed capacity was derived for different scenarios from the USWTDB. The yellow line shows the installed capacity excluding turbines marked as decommissioned. Since most decommissioning dates are not available in the data set, in this scenario, all turbines marked as decommissioned are removed for the complete timespan. The red line indicates the capacity of all installed turbines neglecting decommissioning of turbines. All other solid lines illustrate scenarios under different assumptions of turbine lifetime while ignoring whether the turbines are marked as decommissioned in the database.

    As explained in \cref{sec:appendix-validation-missing_data}, in the USWTDB meta parameters are missing for many turbines. To estimate a capacity for these turbines, mean data imputation for turbines with the same commissioning year was used. The results are shown in solid lines. In contrast, the dashed lines in \cref{fig:irena_capacity_validation} show scenarios, where turbines with unavailable capacity values were simply discarded. Turbines where no commissioning year is available, were removed in all scenarios (see also \cref{sec:methods-data-turbines}).

    In the years before 2005 and after 2010, the USTWDB contains more capacity than the IRENA database when all turbines are included (red line). This could explain the underestimation of system efficiency and output power density using observed data in the years after 2018. Some turbines may have stopped operation after 2018, but where included in the model. This would not change power output using EIA, but overestimate the rotor swept area $A$ and power input $\Pin$ and therefore lead to an underestimation in $\nicefrac{\Pout^\mathrm{EIA}}{A}$ and $\nicefrac{\Pout^\mathrm{EIA}}{\Pin}$. Removing turbines after a certain life time\footnote{Typical wind turbine lifetimes are not below 20 years \cite{ziegler_lifetime_2018}.} could decrease this error, but it would increase the error in earlier years.

    Only in the period 2005--2010, the USTWDB contains less capacity than the IRENA database, when all turbines are included. This could partially explain the overestimation of $\nicefrac{\Pout^\mathrm{EIA}}{A}$ and $\nicefrac{\Pout^\mathrm{EIA}}{\Pin}$ before 2008, but in general, the IRENA database does not indicate, that turbines are missing in the USWTDB.

    For our simulation, we chose the scenario where all turbines are included (red line). This is the scenario with the smallest deviation in output power density using simulated power output $\nicefrac{\mathrm{\Pout}}{A}$ and EIA data $\nicefrac{\Pout^\mathrm{EIA}}{A}$.

\subsection{Missing meta parameters\label{sec:appendix-validation-missing_data}}

\begin{figure}[h]
        \centering{}
        \includegraphics[width=1\textwidth]{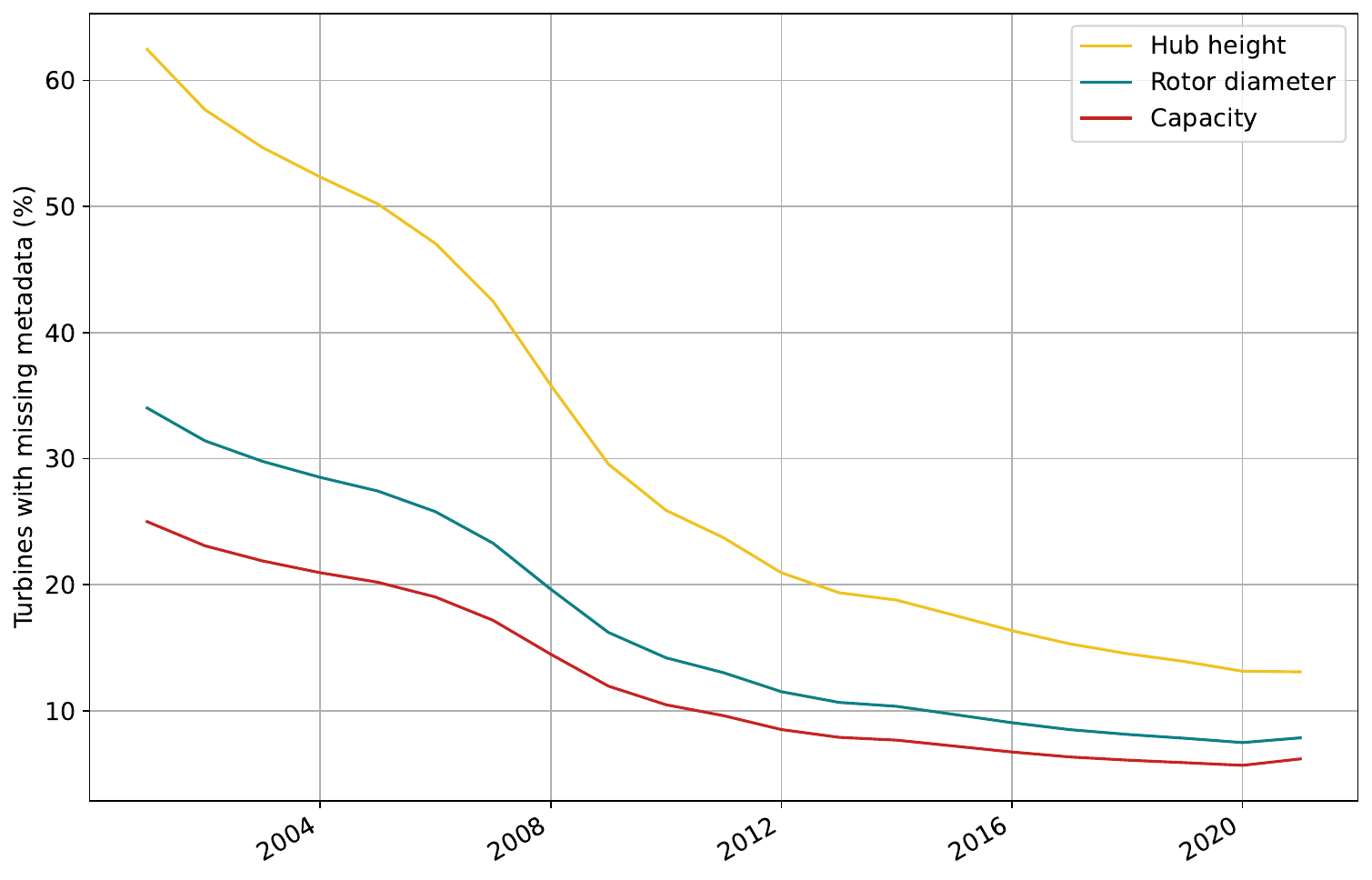}
        \caption{Missing metaparameters in the USWTDB of operating turbines. Decommissioning of turbines is neglected.}
        \label{fig:missing_uswtdb_data}
    \end{figure}

The USWTDB provides a list of all operating wind turbines in the US. However, meta parameters such as rotor diameter, hub height or capacity are missing for many turbines. \cref{fig:missing_uswtdb_data} shows the share of missing meta parameters of turbines installed in a certain year. Missing meta parameters are estimated using mean data imputation for turbines with the same commissioning date. Meta parameters are missing mostly for older turbines. In the period 2001--2008, the share of operating turbines, where no value for capacity is available in the USWTDB, drops from \input{./data/output/data-values/percent_missing_capacity_per_year2001}\unskip\% to \input{./data/output/data-values/percent_missing_capacity_per_year2008}\unskip\%. We assume that this is one of the reasons for the deviation of simulated power output and observed power output in the years before 2008 (see \cref{sec:appendix-validation-poweroutput}).

\subsection{Aging effects\label{sec:appendix-aging-effects}}

Performance of wind turbines decreases over time. Here we use results by \citeauthor{STAFFELL2014775} \cite{STAFFELL2014775} to introduce an aging loss factor to account for losses due to turbine age. \citeauthor{STAFFELL2014775} found that output power decreases 16\% per decade. Hence, we calculated the age $a_{l,Y}$ for each turbine $l$ in each year $Y$ in the period 2001--2021 and then reduce the power output using an aging loss factor, which is linearly scaled to the turbine age, to get an age corrected power output time series $\widehat{\Pout} \left(Y\right)$:

\[
    \widehat{\Pout} \left(Y\right) = \sum_{l\in L} (100\% - 1.6\% \cdot a_{l,Y}) \cdot \sum_{t \in Y} \pout \left(v_{t,l}\right).
\]

\begin{figure}[h]
    \centering{}
    \includegraphics[width=1\textwidth]{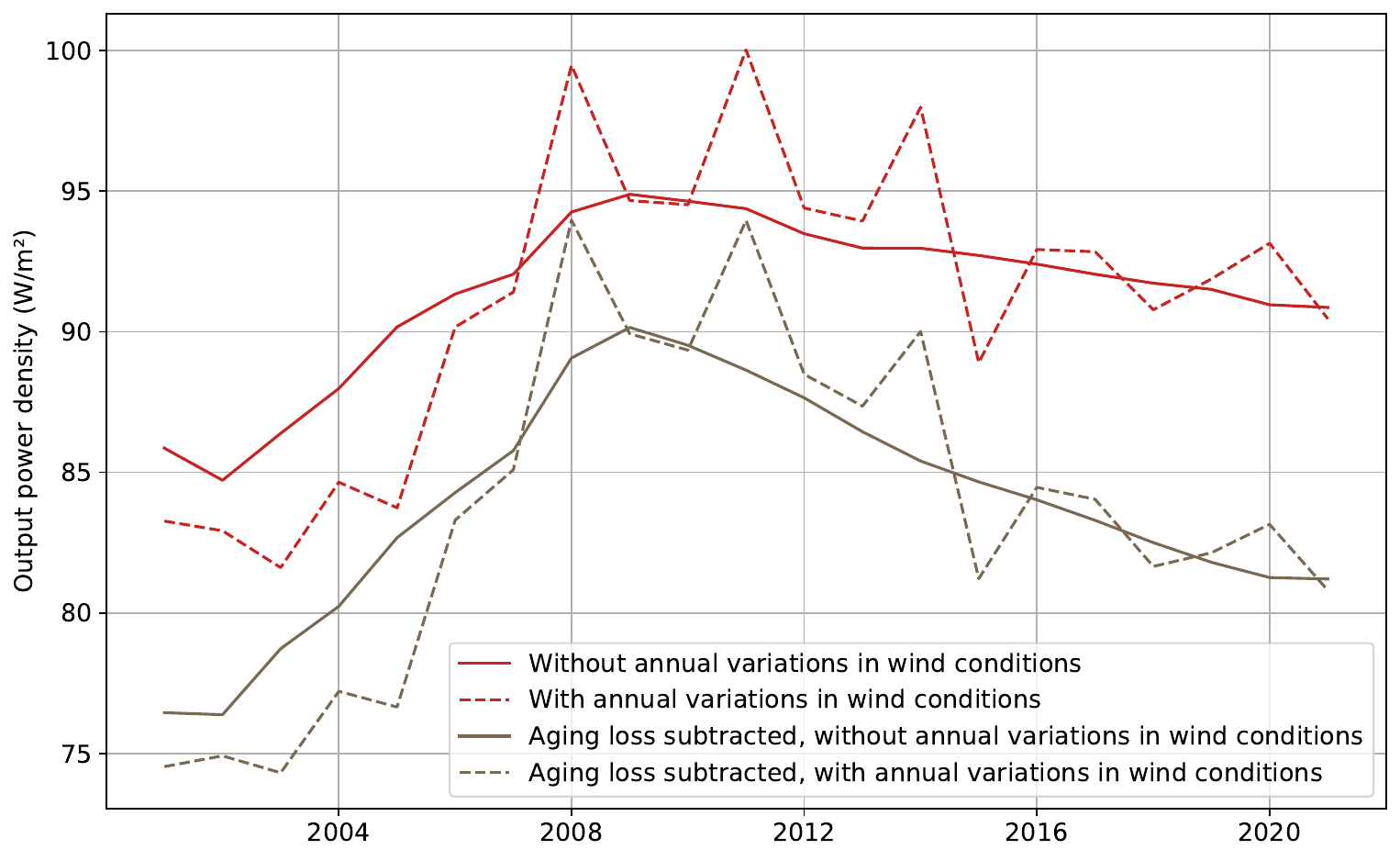}
    \caption{Output power density as used in the main part of the analysis and output power density under the assumption of reduced power output due to turbine aging.}
    \label{fig:d_out_aging}
\end{figure}

\begin{figure}[h]
    \centering{}
    \includegraphics[width=1\textwidth]{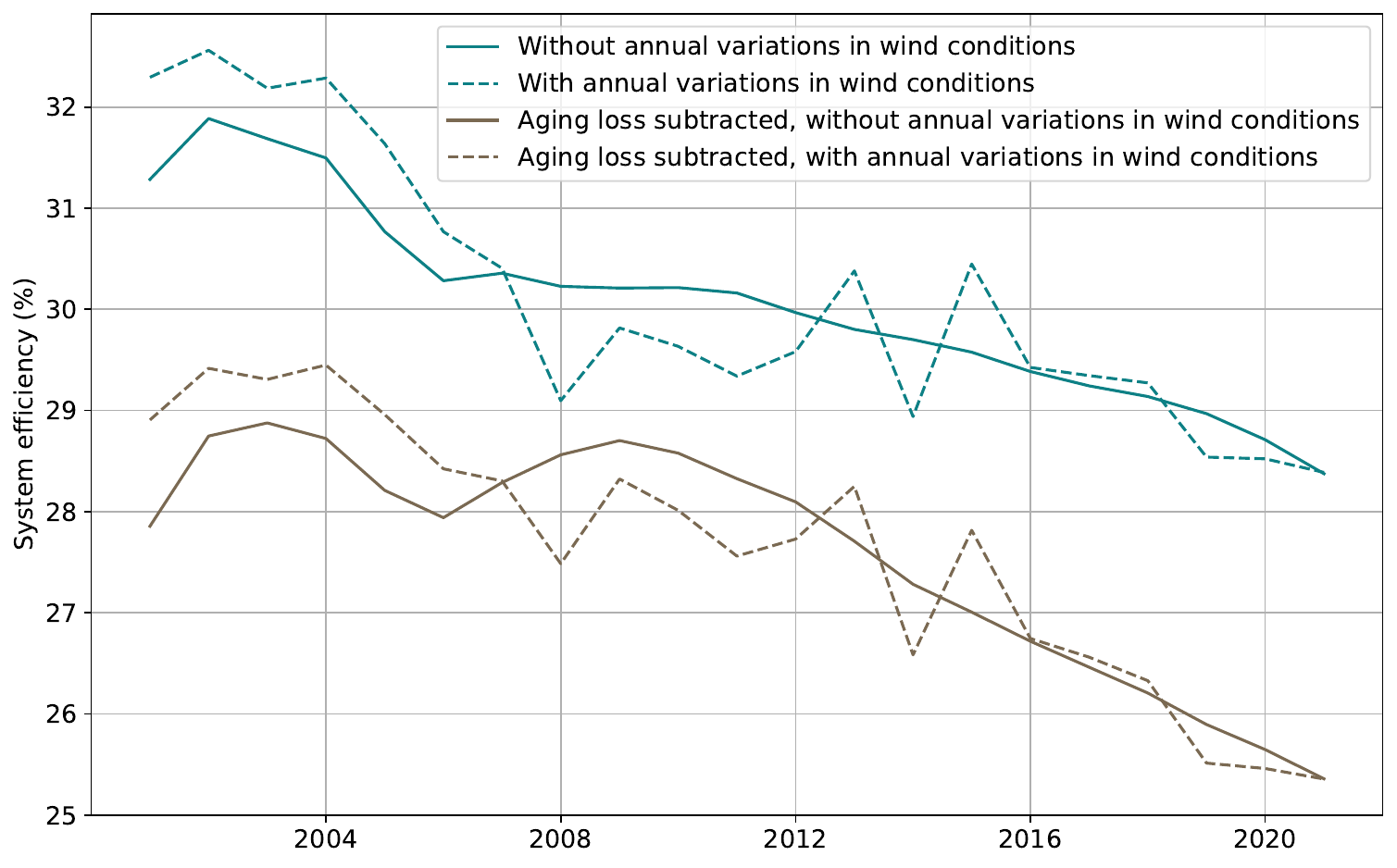}
    \caption{System efficiency as used in the main part of the analysis and system efficiency under the assumption of reduced power output due to turbine aging.}
    \label{fig:efficiency_aging}
\end{figure}

In \cref{fig:d_out_aging} and \cref{fig:efficiency_aging} we compare our results of output power density and system efficiency to results using the age correction factor.

\subsection{Additive decomposition of input power density\label{sec:appendix-additive-decompostion}}

\begin{figure}[h]
        \centering
        \includegraphics[width=1\textwidth]{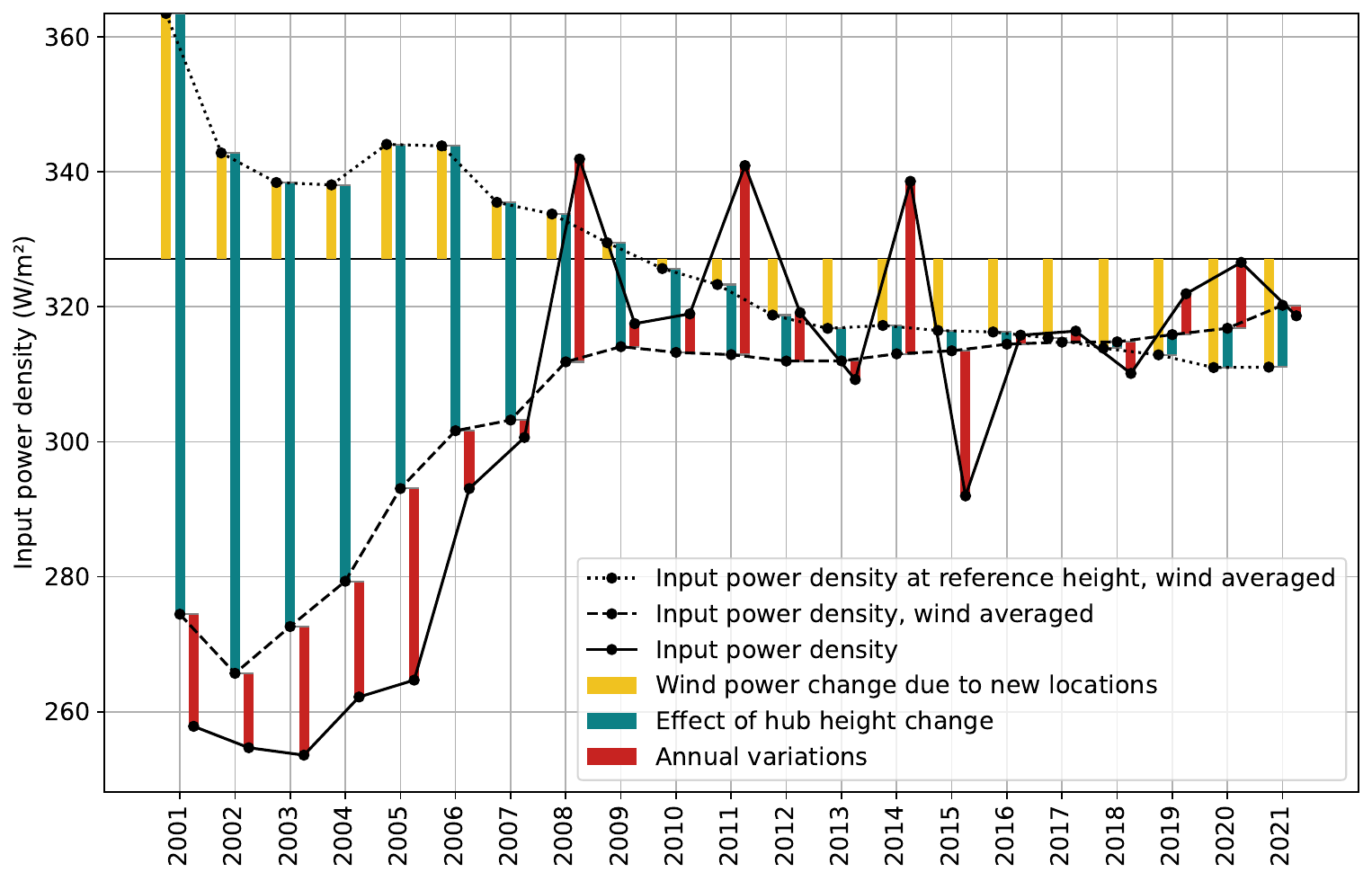}
        \caption{Additive decomposition of input power density.\label{fig:decomposition_pin-waterfall}}
    \end{figure}

    \cref{fig:decomposition_pin-waterfall} illustrates the additive decomposition explained in \cref{sec:results-input-power-density} and in \cref{sec:methods-decomposition-d_in}. The black lines show input power density in different scenarios. The colored bars show the derived effects of wind change due to new locations, of hub height changes and annual variations. The computed input power density is the sum of effects and the base line.

\subsection{Efficiency\label{sec:appendix-efficiency}}

\subsubsection{Side effects of calculations with ratios of aggregated time series\label{sec:appendix-ratio-of-avgs}}

    System efficiency was defined as the ratio of power output and power input $\frac{\Pout}{\Pin}$. Since $\Pin$ and $\Pout$ are temporally and spatially aggregated time series, each data point in the time series of system efficiency is a ratio of averages. System efficiency is not identical with the average coefficient of power
    \[
        \overline{C_\mathrm{p}} = \sum_{l \in L} \sum_{t \in Y} \frac{\pout \left(v_{t,l}\right)}{\pin \left(v_{t,l}\right)},
    \]
    because in general the ratio of averages does not equal the average of ratios. The ratio of averages is a weighted average of ratios, where each summand is weighted by the denominator
    $$
        \frac{\frac{1}{n}\sum_{i=1}^n a_i}
             {\frac{1}{n}\sum_{i=1}^n b_i} =
        \frac{1}{n} \sum_{i=1}^n \frac{b_i}{\frac{1}{n}\sum_{j=1}^n b_j}
        \frac{a_i}{b_i}.
    $$
    This means that hours with high power input contribute more to the aggregated value of system efficiency. Hence, the average coefficient of power $\overline{C_\mathrm{p}}$ and $\frac{\Pout}{\Pin}$ are different measures of wind power efficiency. A comparison is discussed in \cref{sec:appendix-alternative-definitions-efficiency} (see also \cref{fig:average_cp}).

\subsubsection{Correlation between system efficiency and input power density\label{sec:appendix-negative-correlation}}

\begin{figure}[h]
        \centering{}
        \includegraphics[width=1\textwidth]{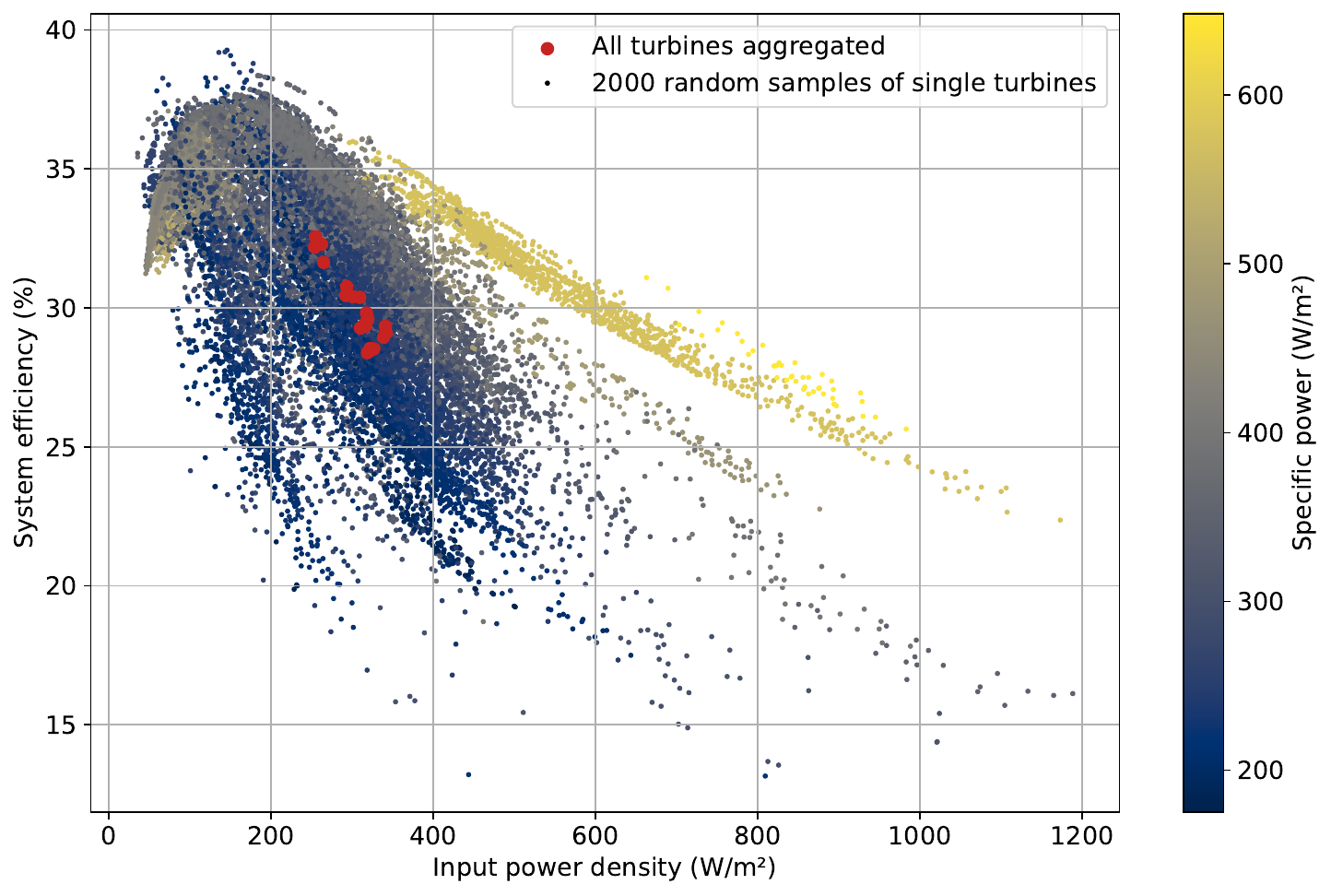}
        \caption{Correlation between system efficiency and input power density of yearly aggregated values time series: one small dot represents one year at one randomly selected (operating) turbine location, one red dot represents one year aggregated over all operating turbines.}
        \label{fig:scatter_efficiency_input_power_density}
    \end{figure}

    Our results show a strong negative correlation between input power density and system efficiency (see \cref{fig:scatter_efficiency_input_power_density}). This is the result of the efficiency of each turbine $C_\mathrm{p}\left(v\right)$ for different wind speeds $v$, the distribution of wind speeds at each turbine location and the spatial and temporal aggregation of $\Pout$, $\Pin$ and $A$.

    The efficiency of a single wind turbine is typically lower at low and high wind speeds. At low wind speeds, there are more losses relative to generated electricity. At high wind speeds, the power output is limited by the size of the generator, i.e. by the nameplate capacity.
    Compared to the curve of $C_\mathrm{p}$, the aggregated efficiency for one turbine is slightly shifted due to the asymmetry of wind speed distributions. In locations with high mean wind speeds, the declining interval of $C_\mathrm{p}$ is dominant and therefore aggregated efficiency is negatively correlated with input power density. Due to the aggregation, locations with high wind speeds have a higher impact on the spatially and temporal aggregated value of system efficiency (see \cref{sec:appendix-ratio-of-avgs}). Therefore system efficiency is negatively correlated to input power density.

\subsubsection{Alternative definitions of efficiency\label{sec:appendix-alternative-definitions-efficiency}}

\begin{figure}[h]
        \centering{}
        \includegraphics[width=1\textwidth]{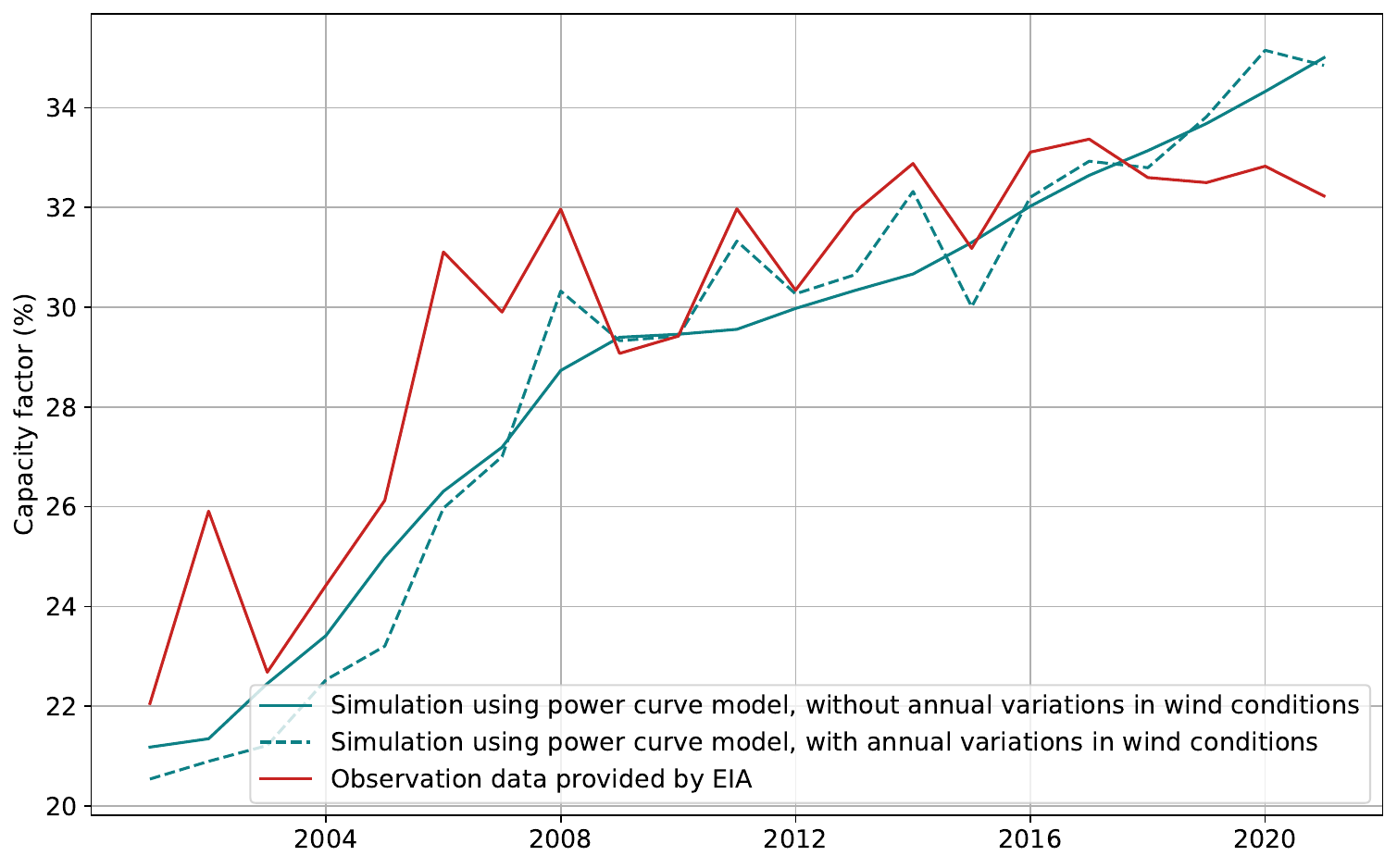}
        \caption{Capacity factors for US wind turbines: ratio of total power output and total installed capacity. Different values for power output are used to calculate capacity factors: simulated values with actual wind conditions, simulated values with long-term average wind conditions and observed values provided by the EIA. Note that this is not the average capacity factor as discussed in \cref{sec:appendix-ratio-of-avgs}. }
        \label{fig:capacity-factors}
    \end{figure}

In general, efficiency is the relation between output and input. We defined system efficiency as ratio between total yearly power output to power input. However, this is not the only relevant measure of wind power efficiency. Instead of using power input, one can also relate generated electricity to the total rotor swept area. Output power density measures how efficiently the available rotor swept area is used. As explained in \cref{sec:wind-power-metrics}, it is the combined effect of the available wind resources (input power density) and the system efficiency. When focusing on land use and its impacts, it makes sense to analyze power output per area of land use. When assuming that the distance between turbines is proportional to their rotor diameters, used land area is also proportional to the total rotor swept area, neglecting effects at the borders of wind parks. Our results of a declining trend in output power density, is therefore in accordance with results by \citeauthor*{miller_corrigendum:_2019} \cite{miller_observation-based_2018,miller_corrigendum:_2019} who show that power output per land use is decreasing (see \cref{fig:d_out_validation}). Specific power is the ratio between capacity and rotor swept area, that means it sets the maximum electricity generation in relation to the available rotor swept area. Opposed to all previously discussed measures of efficiency, which are related to area and wind resources, economic measures of efficiency show an increasing trend. Capacity factors increased in the period 2010--2019 as shown in \cref{fig:capacity-factors}. Since the generator size is one of the most important cost factors when building wind turbines \cite{bolinger_opportunities_2020}, this can be seen as main reason for decreasing LCOE. Hence, when interpreting capacity factors as measure of efficiency, capacity relates to costs and represents the input in the relation between output and input. In contrast, in specific power, capacity stands for the output as it is the maximum power output. Note that LCOE is typically defined as costs per amount of energy, but to make it comparable to other measures of efficiency, the reciprocal of LCOE, energy output per costs, is used here.

    \begin{figure}[h]
        \centering{}
        \includegraphics[width=1\textwidth]{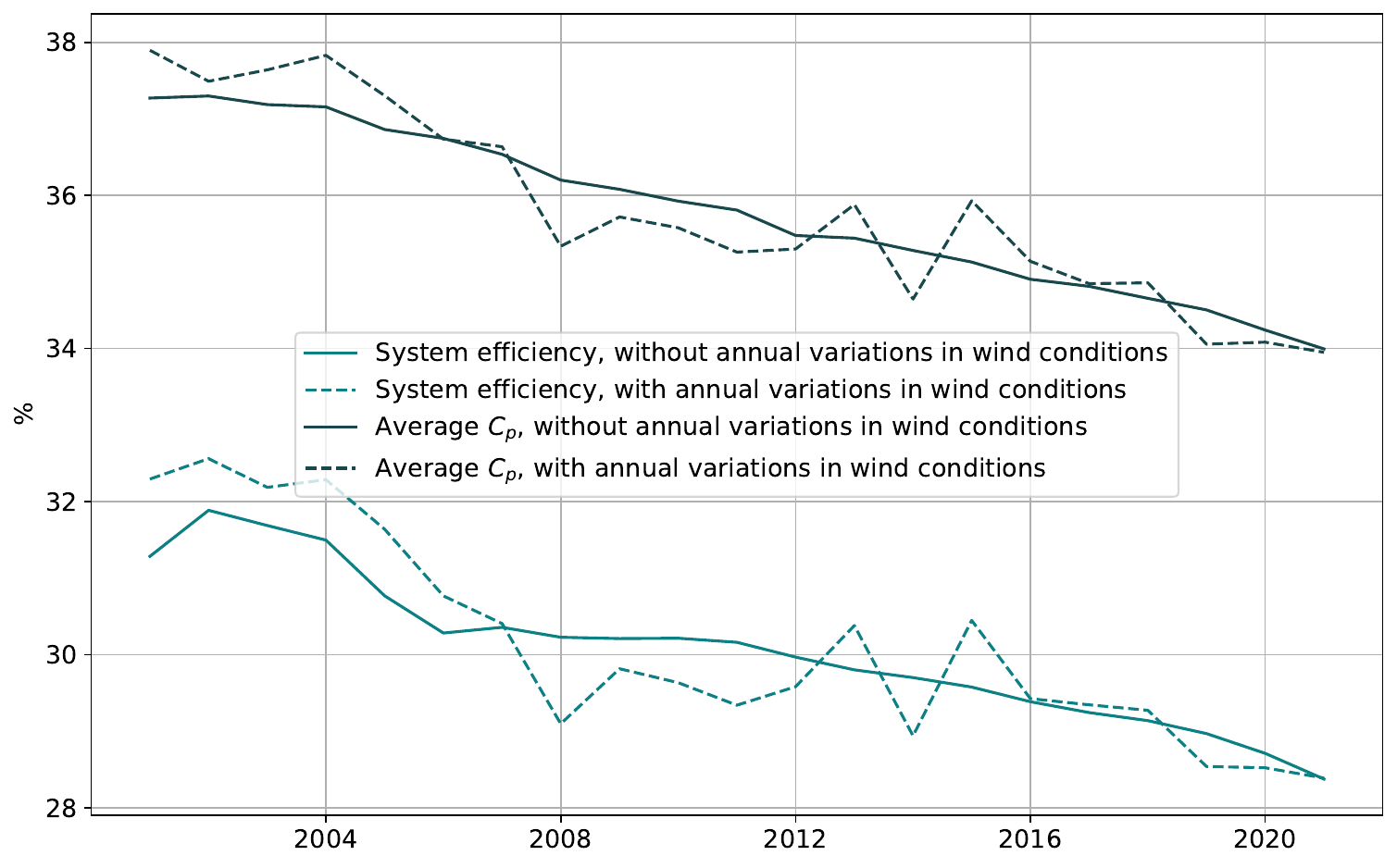}
        \caption{Comparison between system efficiency and average coefficient of power $\overline{C_\mathrm{p}}$ over time: the different weighting of turbines does not change the downwards trend.}
        \label{fig:average_cp}
    \end{figure}

    The average coefficient of power
    \begin{equation}
        \overline{C_\mathrm{p}}\left(Y\right) = \frac{1}{\left|L\right|}\sum_{l\in L} \frac{1}{\left|Y\right|} \sum_{t\in Y} C_\mathrm{p}\left(v_{t,l}\right) =
        \frac{1}{\left|L\right|}\sum_{l\in L} \frac{1}{\left|Y\right|} \sum_{t\in Y} \frac{\pout \left(v_{t,l}\right)}{\pin \left(v_{t,l}\right)}
        \label{eq:average_coefficient_of_power}
    \end{equation}
    is a very similar measure to the system efficiency (\cref{fig:average_cp}). System effects, such as wake effects and maintenance, are not included here. Furthermore, $\overline{C_\mathrm{p}}$ is an average of ratios and not a ratio of averages (see \cref{sec:appendix-ratio-of-avgs}). The average coefficient of power can be interpreted as the average efficiency of an average wind turbine (not weighted by size of the turbine). However, system efficiency can be interpreted as the efficiency of the total wind power production of all installed wind turbines.

    In all discussed definitions of efficiency, measures of produced energy are used. Therefore, the intermittent nature of renewable energy sources is not taken into account. Variability of wind power generation and its correlation with power market prices are of course also important parameters when studying energy systems with high shares of renewables, which are disregarded in the presented efficiency measures.

    \cref{table:alternative-definitions-efficiency} summarizes the discussed alternative measures of efficiency.
    \renewcommand{\arraystretch}{2}
    \begin{table}
        \centering{}
        \rotatebox{90}{
            \begin{tabular}{|m{0.27\textheight}|m{0.25\textwidth}|m{0.14\textheight}|m{0.13\textheight}|m{0.12\textheight}|}
                \hline
                \textbf{Measure of efficiency}                                       & \textbf{Definition}                 & \textbf{Unit} & \textbf{Trend}   & \textbf{Reference}                                              \\ \hline
                System efficiency                                                    & $\frac{\Pout}{\Pin}$                & dimensionless & decreasing       & \cref{sec:results-efficiency}                            \\ \hline
                Output power density                                                 & $\frac{\Pout}{A}$                   & W/m²          & decreasing       & \cref{sec:results-output-power-density}                                       \\ \hline
                Power output per land use                                            & $\frac{\Pout}{A_\mathrm{land}}$     & W/m²          & decreasing       & \cite{miller_observation-based_2018,miller_corrigendum:_2019}   \\ \hline
                Specific power                                                       & $\frac{\textrm{capacity}}{A}$       & W/m²          & decreasing       & \cite{bolinger_opportunities_2020}, \cref{fig:growth_and_specific_power}d                              \\ \hline
                Average coefficient of power $\overline{C_\mathrm{p}}$ & see \cref{eq:average_coefficient_of_power}  & dimensionless & decreasing      & \cref{fig:average_cp}          \\ \hline
                Capacity factors                                                     & $\frac{\Pout}{\textrm{capacity}}$   & dimensionless & increasing       & \cref{fig:capacity-factors}                               \\ \hline
                LCOE (reciprocal)                                                    & $\frac{\Pout}{\textrm{costs}}$      & Wh/\$         & increasing       & \cites{bolinger_opportunities_2020}[58]{irena_renewable_costs}  \\ \hline
            \end{tabular}
        }
        \caption{Overview of alternative definitions of efficiency.}
        \label{table:alternative-definitions-efficiency}
    \end{table}

\subsection{Evolution of turbine characteristics\label{sec:appendix-evolution-turbine-characteristics}}

\begin{figure}
        \centering{}\includegraphics[width=\textwidth]{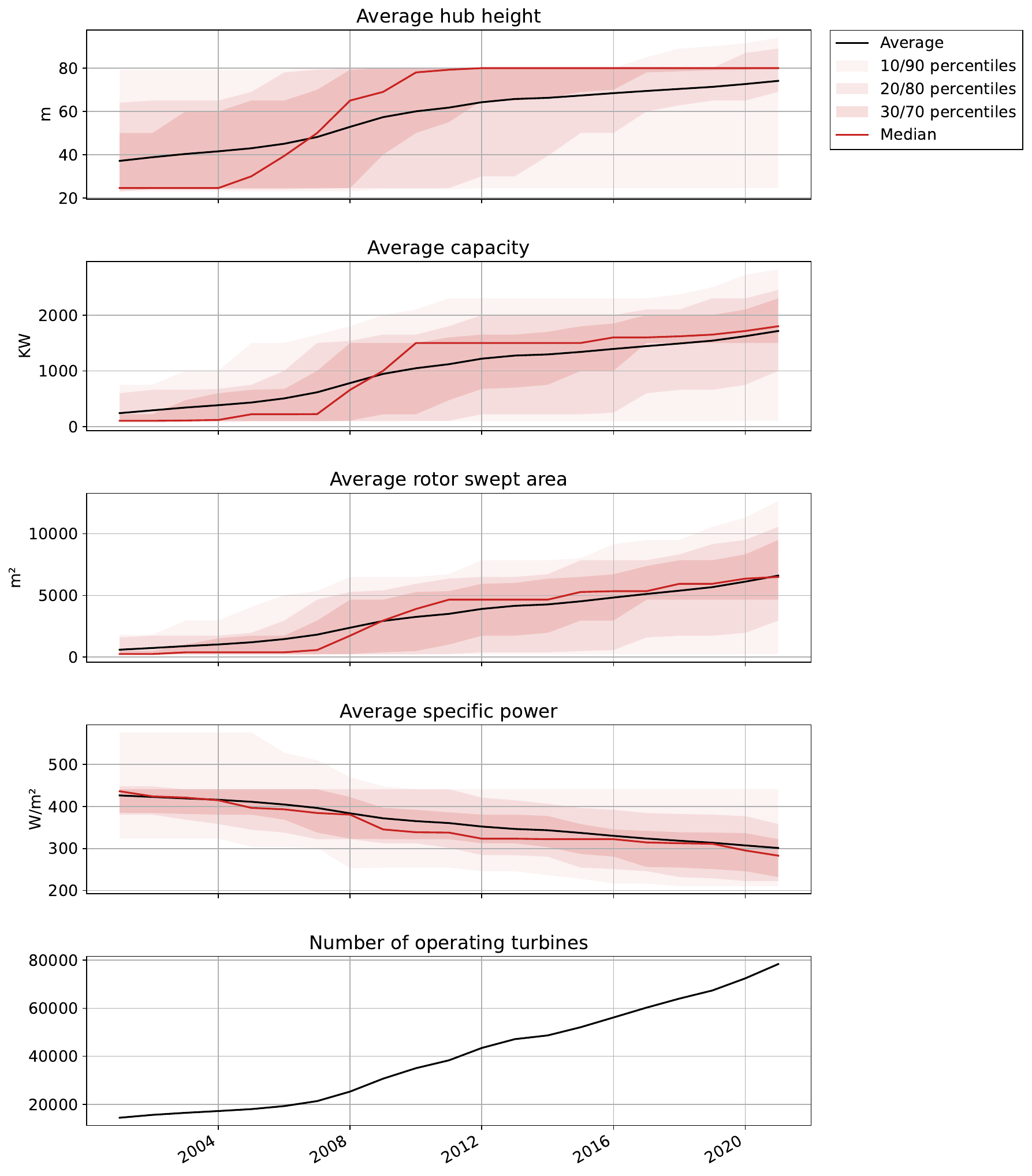}
        \caption{Evolution of turbine characteristics: average hub height, capacity and rotor swept area of operating wind turbine models increases over time, but not at the same pace. Specific power, the ratio between capacity and rotor swept area, shows a declining trend. (Data source: USWTDB, see \cref{sec:methods-data-turbines})
        \label{fig:growth_and_specific_power_with_percentiles}}
    \end{figure}

    \cref{fig:growth_and_specific_power_with_percentiles} shows the evolution of turbine characteristics including the distribution of the variables over all operating turbines in each year (see also \cref{fig:growth_and_specific_power} for a version of the same figure without median and percentiles).

